\documentclass[aps,prb,twocolumn,reprint,superscriptaddress,nofootinbib,preprintnumbers,longbibliography]{revtex4-2}

\usepackage[english]{babel}
\usepackage[utf8]{inputenc}
\usepackage{amssymb}
\usepackage{amsmath}
\usepackage{color}
\usepackage{hyperref}
\usepackage{bbm}
\usepackage{epsfig}
\usepackage{multirow}

\usepackage[normalem]{ulem}

\usepackage[dvipsnames]{xcolor}
\usepackage{xspace}

\newcommand{\e}{\varepsilon}
\newcommand{\s}{\sigma}

\newcommand{\up}{\uparrow}
\newcommand{\down}{\downarrow}
\newcommand{\w}{\omega}

\newcommand{\de}{{\rm d}}

\newcommand{\vk}{\mathbf{k}}

\newcommand{\hc}{\mathrm{h.c.}}

\newcommand{\GF}[1]{\langle\!\langle #1\rangle\!\rangle}

\newcommand{\Aimp}{\mathcal{A}}
\newcommand{\Acb}{\mathcal{B}}

\newcommand{\Ucb}{U_{\rm cb}}

\renewcommand{\Im}{\mathrm{Im}}
\newcommand{\op}[1]{\hat{#1}}

\newcommand{\dk}{\op{d}^\dagger}
\newcommand{\ck}{\op{c}^\dagger}

\newcommand{\da}{\op{d}^{}}
\newcommand{\ca}{\op{c}^{}}

\newcommand{\su}{\mathrm{SU}(2)}
\newcommand{\ru}{\mathrm{U}(1)}
\renewcommand{\S}{\mathcal{S}}
\newcommand{\C}{\mathcal{C}}

\newcommand{\Sec}[1]{Sec.~\ref{sec:#1}}
\newcommand{\beq}{ \begin{equation} } 
\newcommand{\eeq}{ \end{equation} }
\newcommand{\beqa}{\begin{eqnarray}}
\newcommand{\eeqa}{\end{eqnarray}}
\newcommand{\nn}{\nonumber}
\newcommand{\es}{& = &}
\newcommand{\ie}{\textit{i.e.}}
\newcommand{\eg}{\textit{e.g.}}

\newcommand{\fig}[1]{Fig.~\ref{fig:#1}}
\newcommand{\figs}[1]{Figs.~\ref{fig:#1}}
\newcommand{\eq}[1]{Eq.~(\ref{#1})}
\newcommand{\eqs}[2]{Eqs.~(\ref{#1}), (\ref{#2})}

\newcommand{\new}[1]{{\color{ForestGreen}#1}}

\hypersetup{
    bookmarks=false,        
    colorlinks=true,        
    linkcolor=red,        
    citecolor=blue,        
    filecolor=blue,      
    urlcolor=blue
}

\begin{document}

\title{Asymmetry effects on the phases of RKKY-coupled two-impurity Kondo systems}

\author{Krzysztof P. W{\'o}jcik}
\email{kpwojcik@ifmpan.poznan.pl}
\affiliation{Institute of Physics, Maria Curie-Sk\l{}odowska University, 20-031 Lublin, Poland}
\affiliation{Institute of Molecular Physics, Polish Academy of Sciences, 
			 Smoluchowskiego 17, 60-179 Pozna{\'n}, Poland}
\affiliation{Physikalisches Institut, Universit\"{a}t Bonn, Nussallee 12, D-53115 Bonn, Germany}

\author{Johann Kroha}
\email{kroha@th.physik.uni-bonn.de}
\affiliation{Physikalisches Institut, Universit\"{a}t Bonn, Nussallee 12, D-53115 Bonn, Germany}

\date{\today}

\begin{abstract}
  In a related work [\href{https://arxiv.org/abs/2106.07519}{arXiv:2106.07519}]
  we have shown that in the two-impurity Anderson (2iA) model with two hosts
  coupled by spin exchange in the most symmetric case there are either two phase
  transitions or none. The phases comprise the conventional Kondo and RKKY regimes
  and a novel one, interpreted as a Kondo-stabilized, metallic quantum
  spin liquid (QSL).
  Here we analyze how various types of asymmetry affect this picture. We
  demonstrate that the transitions are robust against the coupling and
  particle-hole asymmetries, provided charge transfer is forbidden.
  This holds true despite
  the scattering phase shift at each impurity taking non-universal values.
  Finally, for an extended model including charge transfer between the hosts 
  and a small Coulomb interaction at the host sites directly coupled to impurities, 
  we show that the presence of charge transfer changes the phase transitions 
  into crossovers. Provided the inter-host hopping is sufficiently small, 
  this leads to qualitatively the same physics at non-zero temperature. 
  The relevance of this model for rare-earth atoms in a metallic host 
  is discussed and potential experimental setups for observing our findings 
  are proposed.
\end{abstract}

\maketitle

%%%%%%%%%%%%%%%%%%%%%%%%%%%%%%%%%%%%%%%%
%%%%%%%%%%%%%%%%%%%%%%%%%%%%%%%%%%%%%%%%
\section{Introduction}
\label{sec:intro}
%%%%%%%%%%%%%%%%%%%%%%%%%%%%%%%%%%%%%%%%
%%%%%%%%%%%%%%%%%%%%%%%%%%%%%%%%%%%%%%%%

Heavy fermion (HF) materials have been studied for many years,
yet their rich phase diagrams still elude precise 
understanding \cite{Coleman_FrustrationAndKondoInHF,Kirchner2020Mar}.
In particular, already in 1977 Doniach 
proposed a scenario \cite{Doniach} for HF magnetic phase transitions  
to be qualitatively captured by the competition of the local 
Kondo screening of individual impurities and the long-range
conduction-band-mediated, indirect spin exchange between 
distinct impurities, the RKKY interaction \cite{RK,K,Y}. 
This local perspective gained particular interest after it 
has been shown that already in the case of two impurities it can trigger 
a quantum phase transition (QPT) \cite{Jones1,Jones2,Jones3},
related to two-channel Kondo physics \cite{Gan1995Mar,Mitchell2012Feb}.
Since then, two-impurity or dimer systems have become 
a test ground for many concepts concerning HF properties 
\cite{Bork,Bayat2014May,Pruser2014Nov,Spinelli2015Nov,Moro-Lagares2019May}.
This strategy is further supported by the recent, successful
dynamical mean-field theory (DMFT) 
mapping of a generic Kondo lattice onto self-consistent
two-impurity problem \cite{Gleis}.

Nevertheless, the significance of the Jones-Varma QPT for HF 
criticality is still debated for a number of reasons.\\ 
(1) It relies on strong assumptions concerning, 
in particular, a special type of particle-hole (PH) symmetry,
and the QPT is in general smeared into crossover in less 
symmetric cases \cite{Fye,Affleck,Silva1996Jan}. In real materials
PH symmetry is usually not present. Still, a properly tuned 
counter-term allows to restore the transition in the partly 
symmetric model \cite{Fabian}. The absence of charge transfer
between the hosts suffices to observe the transition in theory
for two impurities \cite{Affleck,Zarand2006Oct},
and self-consistency restores the stability of the transition 
within DMFT of the Kondo lattice with anti-ferromagnetic 
order \cite{Gleis}. 
In experiments, even the presence of a small charge transfer seems
to not suppress the two-impurity QPT \cite{Bork}.\\
(2) In a dense impurity lattice there may 
be too few electrons for complete screening of all the impurities 
independently, as suggested by Nozieres' exhaustion principle
\cite{Nozieres1985,Nozieres1998}
and proven recently for the periodic Anderson model \cite{Eickhoff2020Nov}.
This suggests a reinterpretation of the metallic HF system as  
composed of partially correlated impurities 
instead of individually Kondo-screened ones. 

(3) It has been stressed that in many HF materials
an important role is played by frustration \cite{Si2010,Coleman_FrustrationAndKondoInHF} which competes with ordering tendencies and may lead
to exotic phases, including metallic QSLs
\cite{Nakatsuji2006Mar,Friedemann2009Jul,Lucas2017Mar,Zhao2019Dec,Majumder2022May,Tripathi2022Aug},
coexistence of the Kondo effect and QSL \cite{Nobukane2020Jan}
or magnetic ordering of residual local magnetic moments 
\cite{Lacroix1996Dec,Li2010Mar,Motome2010Jul,Bernhard2015Sep,Sato2018Mar,KWIWJK_3QD,Kessler2020Dec}.
This intensively researched field applies to materials possessing magnetic
moments arranged into triangular or other frustrated lattices,
but seems less related to cases lacking geometrical frustration,
such as studied here. 

(4) The Jones-Varma model neglects that the realistic
  RKKY interaction is conduction-electron mediated, but replaces it by
  a direct Heisenberg exchange between the impurities as an
  independent parameter.
The resulting phase diagram is different from the 2iA model with an
individual host for each impurity and a spin exchange between the two
hosts \cite{KWJK_2imp}. Remarkably, there a QSL phase occurs  
even without geometrical frustration, but stabilized against a dimerized
phase by the Kondo effect \cite{KWJK_2imp,Coleman1989Jul,Andrei1989Jan}.
Most recently, the RKKY coupling is also considered as an indirect 
interaction in Ref.~\cite{Peschke2022Sep} for a Kondo necklace model, 
but there the lack of charge degrees of freedom excludes a QSL phase 
studied here.

In the present paper, we take Ref.~\cite{KWJK_2imp} as a starting 
point and analyze how the results presented there depend on different 
types of asymmetry.
In \Sec{model} we introduce the model, explaining different types of 
asymmetry in Secs.~\ref{sec:modelSSS}-\ref{sec:modelSSZ} and the 
methodology in \Sec{NRG}. 
The simplicity of the proposed two-impurity setup allows for a near-exact
numerical solution, in particular without any type of mean-field treatment
nor perturbative approximations.
In \Sec{sym} we give detailed definitions of all crossover scales
relevant to the model, 
and elaborate on the role of interactions in the conduction band 
for the stability of the spin liquid phase.
\Sec{results} is devoted to the description of the results
in different asymmetric cases. 
In particular, we show that the zero-temperature phase diagram obtained
in Ref.~\cite{KWJK_2imp} stays intact despite PH or Kondo coupling
asymmetry.
Then, in \Sec{SS}, we show that the presence of charge transfer changes 
the QPTs into crossovers, such that soft-boundary regimes of similar 
properties replace the well-defined phases of the symmetric model.
We conclude in \Sec{con}.

%%%%%%%%%%%%%%%%%%%%%%%%%%%%%%%%%%%%%%%%
%%%%%%%%%%%%%%%%%%%%%%%%%%%%%%%%%%%%%%%%
\section{Model and its symmetries}
\label{sec:model}
%%%%%%%%%%%%%%%%%%%%%%%%%%%%%%%%%%%%%%%%
%%%%%%%%%%%%%%%%%%%%%%%%%%%%%%%%%%%%%%%%

\subsection{Fully symmetric case}
\label{sec:modelSym}

\begin{figure}[tb]
\includegraphics[width=\linewidth]{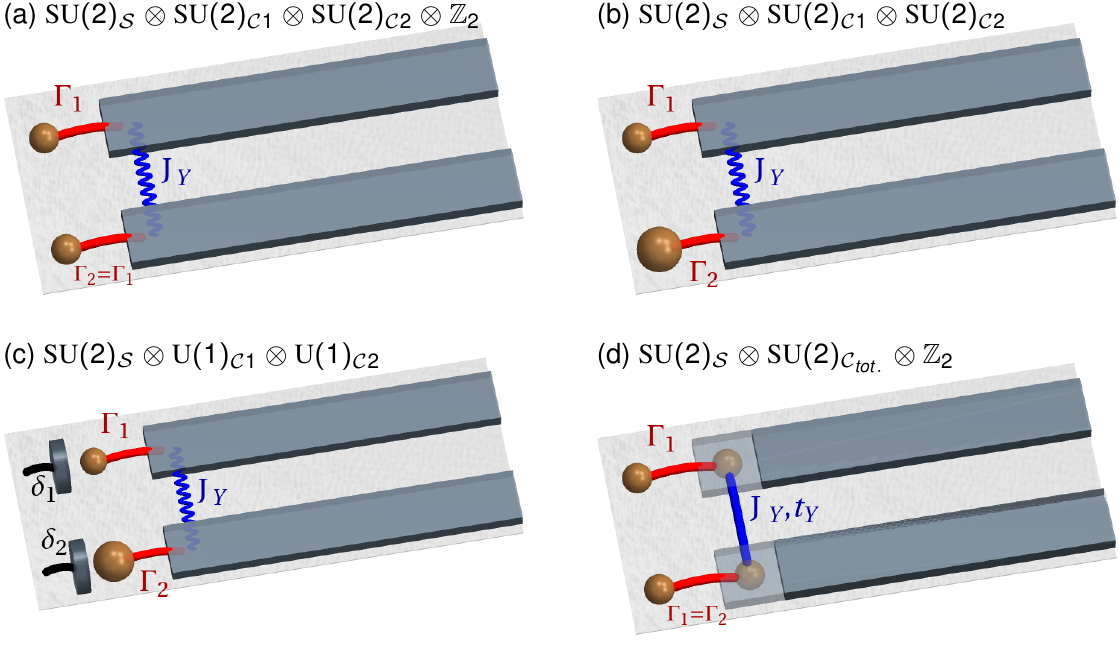}
\caption{(a) Schematic illustration of the fully symmetric model of Ref.~\cite{KWJK_2imp}. 
		 (b-d) Different types of asymmetric models considered in the present paper:
		 (b) model with channel asymmetry,
		 (c) model with PH asymmetry,
		 (d) model with allowed inter-channel charge transfer.
		}
\label{fig:models}
\end{figure}

The model consists of two impurities, each one coupled to a different host.
The hosts are coupled by a spin exchange, as depicted schematically 
in \fig{models}(a).
The Hamiltonian is based on the Anderson impurity model for each of
the \emph{channels}, where each host together with 
its corresponding impurity is referred to as a channel. 
Therefore, its symmetrical form reads
\begin{eqnarray}
H &&=
	\sum_{\alpha\vk\s} \e_{\alpha\vk} \ck_{\alpha\vk\s} \ca_{\alpha\vk\s} 
	+ \sum_{\alpha\vk\s} V_{\alpha} (\ck_{\alpha\vk\s} \da_{\alpha\s} +\hc)
\nn\\&&\;\;\;
    + \sum_{\alpha\s} \e_\alpha \hat{n}_{\alpha\s}
	+ U \sum_{\alpha} \hat{n}_{\alpha\up} \hat{n}_{\alpha\down} 
	+ J_Y \vec{\hat{s}}_1 \dot \vec{\hat{s}}_2, \quad
\label{Hsym}
\end{eqnarray}
where $\hat{n}_{\alpha\s} = \dk_{\alpha\s}\da_{\alpha\s}$ is the number operator 
of spin-$\s$ electrons on the impurity $\alpha$ ($\alpha\in\{1,2\}$ and $\s\in\{\up,\down\}$),
the conduction-band spin operator at the impurity site in channel $\alpha$ is defined as 
$\vec{\hat{s}}_{\alpha} = \sum_{\vk\vk'} \ck_{\alpha\vk\s} \vec{\s}_{\s\s'} \ca_{\alpha\vk'\s'}$, with $\vec{\s}$ the vector of Pauli matrices. $V_1=V_2 \in \mathbb{R}$ determine 
the hybridization between the impurity and the host in respective channel. 
Finally, $U$ denotes the Coulomb repulsion within the impurity orbitals,
while $J_Y$ is the inter-host spin-exchange coupling.
$J_Y$ with suppressed charge transport may possibly be created by
large-spin molecules or a chain of magnetic atoms in between the two
impurities, as in Ref.~\cite{Zarand2006Oct}. The impurity-host hybridization functions are $\varGamma_{\alpha\s}(\w) = \pi \rho_{\alpha\s}(\w) V_{\alpha}^2$, 
where $\rho_{\alpha\s}$ denotes the density of host states. 
We assume a constant, spin-independent and PH symmetric host density of
states within the bandwidth $D_{\alpha}$,
$\rho_{\alpha\s}(\w) \equiv N^{-1} \sum_{\vk} \delta(\w - \e_{\alpha\vk\s}) \approx (2D_\alpha)^{-1}$ for $|\w| \leq D_\alpha$ ($N$ is the number of $\vk$ points in momentum space) and $\rho_{\alpha\s}(\w)=0$ for $|\w| > D_\alpha$.
We will comment on the case $D_1 \neq D_2$ when discussing channel asymmetry, see \Sec{SSS}, and assume $D_1=D_2=D$ otherwise. 
As long as $\rho_{\alpha\s}(\w)$ is regular at the Fermi level,
their energy dependence is expected to be unimportant at low temperatures \cite{WilsonNRG}. Discussing the significance of (partial) spin polarization of the leads would be relevant for ferromagnetically correlated leads and require an additional study.

Setting the energy levels of the impurities at 
$\e_\alpha = -U/2$ ensures the full PH symmetry
of the model, while independence of the Hamiltonian 
parameters of $\alpha$ guarantees channel symmetry.
The symmetry of the model is then the same as the one  
studied originally by Jones and Varma \cite{Jones2}. 
The total symmetry of the model is a product of several subgroups, 
$\su_{\S}\otimes \su_{\C_1}\otimes \su_{\C_2}\otimes \mathbb{Z}_2$,
described as follows.
The first term corresponds to conventional total spin (denoted $\S$) conservation in the absence of anisotropy. Furthermore, the electric charge is conserved in each of the channels separately, and in the presence of PH symmetry the conventional $\ru_{\C_\alpha}$ charge symmetries 
are lifted to an isospin $\su_{\C_\alpha}$, where the $z$-component of the isospin is the physical charge, and the isospin rising and lowering corresponds
to PH transformations.
Finally, the $\mathbb{Z}_2$ symmetry corresponds to the invariance of the
Hamiltonian with respect to interchanging the channels.

Noteworthy, in the conventional 2iA model with a single host, the
$\su_{\C\alpha}$ charge symmetries do in general not appear for both channels
of the effective NRG model separately \cite{Fabian}. The different types of
asymmetry taken into consideration in the present paper are explained
one by one in the following and examined for their influence on the
phase diagram of the model in \Sec{results}.

\subsection{Channel asymmetry}
\label{sec:modelSSS}

Channel asymmetry appears when the hybridizations of impurities with their 
respective hosts differ between the channels, \ie{} $\varGamma_1 \neq \varGamma_2$
as in \fig{models}(b), or $V_1\neq V_2$ in \eq{Hsym}. 
This, in turn, implies different Kondo 
couplings and different Kondo temperatures,
$T_{{\rm K}1}$ and $T_{{\rm K}2}$, characterizing the corresponding channels.
Due to the exponential dependence of $T_{{\rm K}\alpha}$ on $\varGamma_\alpha$,
even a small difference in $\varGamma_\alpha$ renders large discrepancy
of Kondo scales. Hence, the discussion of this asymmetry seems necessary
for making a reliable connection to experimental 
reality. Technically, this means that the $\mathbb{Z}_2$ symmetry is
broken, while all three $\su$ symmetries remain intact.

\subsection{Particle-hole asymmetry}
\label{sec:modelSUU}

The PH symmetry plays a special role in the context 
of the Jones-Varma two-impurity model. In fact, it has 
been recognized that only a special type of PH symmetry 
guarantees the existence of QPT which is otherwise turned
into a crossover \cite{Fye,Affleck}. However, it has subsequently 
been proven that the QPT is robust even in the absence of
PH symmetry and destabilized only by charge transfer
between the hosts \cite{SelaAffleck,Zarand2006Oct}.
Moreover, the only marginally 
relevant perturbation around the QPT fixed point can be 
eliminated by an appropriate counter-term, such as 
additional inter-impurity hopping \cite{Fabian}. Therefore, 
it seems that the analysis cannot be complete without checking
if the Kondo-RKKY transition and the QSL 
transition, reported in Ref.~\cite{KWJK_2imp}, share
this characteristics. 
To this end, we set $\e_\alpha = -U/2 + \delta_\alpha$ 
in \eq{Hsym} and study the properties of the system for
different detunings from the PH symmetry point $\delta_\alpha$,
as schematically presented in \fig{models}(c). 
$\delta_\alpha \neq 0$ reduces the charge symmetry of channel $\alpha$
from $\su$ to regular $\ru$ charge conservation.

As a particular case, PH asymmetry includes the situation when 
individual channels are asymmetric, yet compensate each 
other to restore global (weak) PH symmetry, $\delta_1 = -\delta_2$
for $\varGamma_1 = \varGamma_2$. However, as shown below,
also in the more general case of independent detuning 
of the energy levels of each impurity the transitions stay 
intact.

\subsection{Charge transfer}
\label{sec:modelSSZ}

We also analyze the situation of allowed charge 
transfer between the hosts. By this we mean adding a
hopping term $t_Y$ to  \eq{Hsym}, that is, considering the
Hamiltonian
\begin{eqnarray}
\label{Hasym}
H_{\rm asym} &=&
	H
	+ \Ucb \sum_{\alpha} \ck_{\alpha\up}\ca_{\alpha\up}\ck_{\alpha\down}\ca_{\alpha\down}
	\nn\\&&
	+ t_Y \sum_{\s} ( \ck_{1\s}\ca_{2\s} + \hc )
	, \quad
\end{eqnarray}
where $\ca_{\alpha\s} = \sum_{\vk} \ca_{\alpha\vk\s}$ denotes the
conduction-electron operator with spin $\s$ at the site of the
impurity $\alpha$.
Note, that in this case the two hosts sites are, in fact,  
interacting, as schematically depicted in \fig{models}(d).

Introducing $\Ucb > 0$ is well motivated in particular in the 
case when the impurities are $f$ or $d$-electron ad-atoms,
such as Ce or Co, on a metal surface. 
Then, the $f$ or $d$-electron carries the Kondo impurity spin
and simultaneously hybridizes with the $s$-electrons of the
same impurity atom, 
where the local Coulomb interaction is much weaker, $\Ucb < U$.
In turn, the $s$-orbital is coupled to the itinerant host electrons \cite{Bork}.
We note in passing that a similar model with $U_{\rm cb} \neq 0$ would 
emerge in DMFT for an anti-ferromagnetic
phase of the Kondo-Hubbard model relevant for manganites, even though 
the Kondo coupling would need to be replaced by ferromagnetic 
Hund's exchange there \cite{Held2000May,Hafez-Torbati2021Dec}.

Since asymmetries different from charge transfer terms are 
irrelevant (see above), below we will restrict ourselves
to the PH- and channel-symmetric case for the sake of simplicity.
Nevertheless, the results are expected to apply to the general 
case.

\subsection{Methods}
\label{sec:NRG}

The model is solved by the numerical renormalization 
group (NRG) procedure \cite{WilsonNRG,NRG_RMP}. Our implementation is based 
on the open-access code Ref.~\cite{fnrg}, which uses the  basis set 
of all discarded states \cite{AndersSchiller1} to construct the
full density matrix of the system \cite{Weichselbaum}. 
This allows to calculate static expectation values for any temperature $T$
and compute arbitrary spectral functions in a sum-rule-conserving  
framework, directly in their Lehmann representation. 
In all NRG calculations we use the NRG discretization
parameter $\Lambda=2.5$. At each NRG step we keep all states with
(rescaled) energy $E_j$ below a cutoff value $E_{\rm cut}$,
where $E_{\rm cut}$ is chosen within the range $6.5 < E_{\rm cut} < 7.0$ (in units of the iteration scale) such that the energy difference between the last
kept and the first discarded state is greater than $0.001$.
For all calculations we choose Hamiltonian parameter values (as given 
in the following) such that the intrinsic energy scales are well separated 
and the different regimes can be clearly identified.

To avoid ambiguities introduced by artificial broadening of 
the spectral functions, instead of studying the spectral densities 
directly, we will use the impurity conductance $G$ and the conduction-band 
conductance $g$, defined as
\beqa
G_{\alpha}(T) \es \sum_\s\int \Aimp_{\alpha}(\w) \left(-\frac{\partial f_T(\w)}{\partial\w}\right) \de\w , 
\label{G} \\
g_{\alpha}(T) \es \sum_\s\int \Acb_{\alpha}(\w) \left(-\frac{\partial f_T(\w)}{\partial\w}\right) \de\w , 
\label{g}
\eeqa
where $\Aimp_{\alpha} = - \varGamma_\alpha \Im \GF{\da_{\alpha\s} ; \dk_{\alpha\s}}^{\rm ret}(\w)$ 
is the normalized spectral density at the impurity $\alpha$,
which is independent of $\s$ due to $\su$ spin symmetry, 
and similarly $\Acb_{\alpha} = -2D_\alpha \Im \GF{\ca_{\alpha\s} ; \ck_{\alpha\s}}^{\rm ret}(\w)$.
$f_T(\w)$ denotes the Fermi-Dirac distribution function at temperature $T$
and $\GF{\ldots}^{\rm ret}$ the retarded fermionic Green function in frequency space.
With the definitions \eq{G} and (\ref{g}), the conductance measured with an STM tip 
coupled to impurity $\alpha$ (conduction band site in direct vicinity of impurity $\alpha$) 
is proportional to $G_\alpha$ ($g_\alpha$), respectively. Since the actual conductance 
would depend on the coupling strength between the STM and the impurity (host), we normalize
it by its maximal value $G_0$ ($g_0$), corresponding to resonant
transport conditions.

%%%%%%%%%%%%%%%%%%%%%%%%%%%%%%%%%%%%%%%%
%%%%%%%%%%%%%%%%%%%%%%%%%%%%%%%%%%%%%%%%
\section{Crossover temperatures and properties of the spin-liquid phase}
\label{sec:sym}
%%%%%%%%%%%%%%%%%%%%%%%%%%%%%%%%%%%%%%%%
%%%%%%%%%%%%%%%%%%%%%%%%%%%%%%%%%%%%%%%%

In this section we corroborate and extend the results of Ref.~\cite{KWJK_2imp} in that we provide a detailed explanation of the relevant energy scales, 
an analysis of the fixed point spectra in the respective
phases of the model, and a more extensive discussion of the Heisenberg
transition in the conduction band. We consider the symmetric model
of \eq{Hsym}, with $D_1=D_2=D=2\,U$ and $\varGamma_1=\varGamma_2=\varGamma =0.0488\,U$,
which results in a single-impurity ($J_Y=0$) Kondo temperature of
$T_{{\rm K}\alpha}^0 \approx 10^{-4}\,U$. It exhibits two QPTs when $J_Y$
is increased from $0$ to $J_Y \gg D$ \cite{KWJK_2imp}. These results will
then be used as a reference for the asymmetry analysis in \Sec{results}.

\begin{figure}
\includegraphics[width=0.9\linewidth]{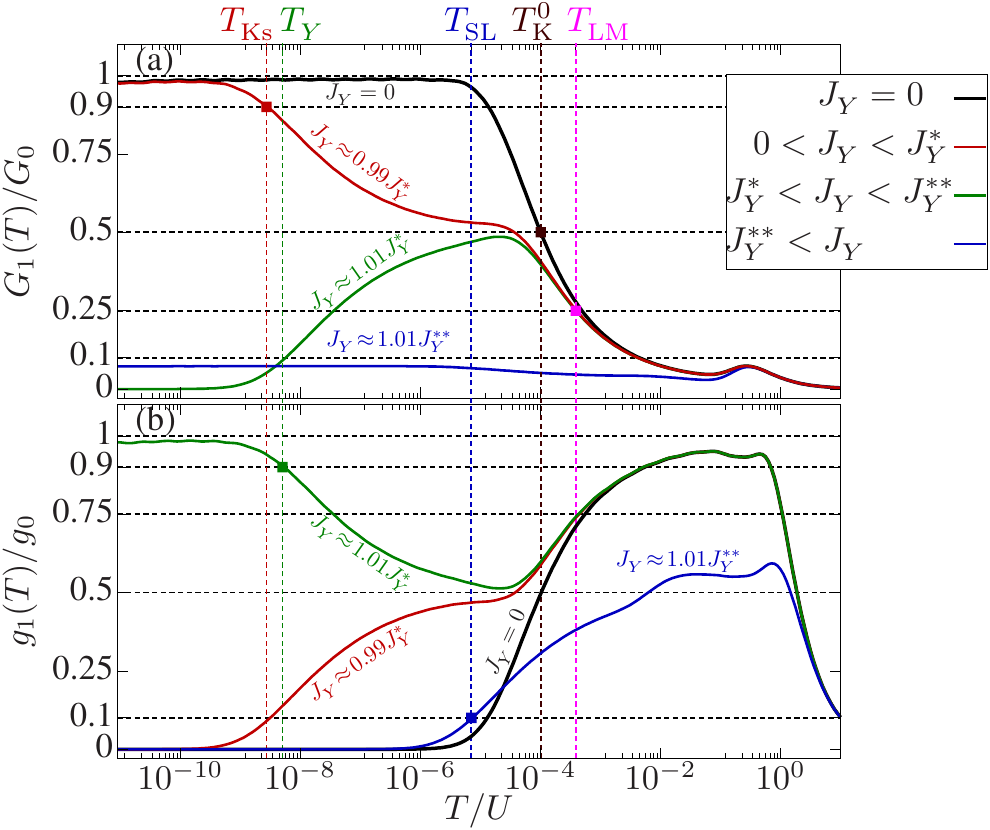}
\caption{(a) Impurity conductance $G_1$ and (b) conduction-band 
		 conductance $g_1$ as functions of temperature $T$ 
		 (note the logarithmic scale) for a symmetric model 
		 with $\varGamma_1 = \varGamma_2 = 0.0488 U$
		 ($T_{{\rm K}1}^0 = T_{{\rm K}2}^0 \approx 10^{-4}U$),
		 and a few values of $J_Y$, representative for various regimes
		 (the actual values are $J_Y/D \in\{ 0, 0.105, 0.107, 1.5 \}$).
		 Other parameters: $D=2U$, $\Lambda=2.5$, $6.5 < E_{\rm cut} < 7$.
		 }
\label{fig:sym}
\end{figure}

\subsection{Crossover temperatures and Kondo destruction}

We first introduce a number of crossover scales that characterize the NRG flow of the fully symmetric system from the weak-coupling local-moment fixed point at high energy to the low-energy fixed points describing the various ground-state phases of our system, the Kondo, the RKKY, and the QSL phase, respectively, depending on the value of $J_Y$. This flow can be observed in the $T$ dependence of the $G_1$ and $g_1$ as shown for the symmetric case ($G_2=G_1$ and $g_2=g_1$) in \fig{sym} (a), (b). Above the conduction bandwidth, $T>D=2U$, $G_1$ and $g_1$ approach zero due to the absence of spectral density. $G_1$ features a bump at $T\approx U/2$ where excitations to empty and doubly occupied impurity states are thermally accessible. Correspondingly, $g_1$ has a dip at $T\approx U/2$ due to a Fano-like depletion of conduction spectral density. It is evident from the figure that in the general case ($J_Y\neq 0$), for $T<U/2$ the temperature dependence is governed by two scales. In the local-moment regime (decoupled impurity spins) we have $G_1/G_0\approx 0$ and $g_1/g_0\approx 1$, so that we can define the scale below which the system deviates from the local-moment regime as the temperature $T_{\rm LM}$ where $G_1(T_{\rm LM})/G_0=0.25$ and $g_1(T_{\rm LM})/g_0=0.75$, see \fig{sym} (a), (b).

The low-$T$ scales can be read off from \fig{sym} as follows. 
The Kondo regime is characterized by $G_1(0)/G_0 = 1$, $g_1(0)/g_0 = 0$, while in the RKKY regime $G_1(0)/G_0 = 0$, $g_1(0)/g_0 = 1$. Consequently, we can define the scale on which the Kondo fixed point is approached ($0<J_Y<J_Y^*$) as the strong-coupling Kondo temperature $T_{\rm Ks}$  where $G_1(T_{\rm Ks})/G_0=0.9$ and $g_1(T_{\rm Ks})/g_0=0.1$ (red curves), while the approach to the RKKY fixed point ($J_Y^*<J_Y<J_Y^{**}$) is characterized by the scale $T_Y$ with $G_1(T_Y)/G_0=0.1$, $g_1(T_{Y})/g_0=0.9$ (green curves). The approach to the QSL fixed point ($J_Y>J_Y^{**}$) is more difficult to characterize as the impurity conductance $G_1$ assumes non-universal low-$T$ values in this case [blue curve in \fig{sym} (a)], while $g_1(0)/g_0=0$. Therefore, we define the QSL scale $T_{\rm SL}$ as the temperature where $g_1(T_{\rm SL})=0.1$ [blue curve in \fig{sym} (b)].

We note that in the limit of a single Anderson impurity ($J_Y=0$) the strong and weak coupling scales $T_{\rm Ks}$ and $T_{\rm LM}$ become proportional to each other (not shown), signaling the universality of the single-impurity Anderson model. We find that the Kondo fixed point is approached at low $T$ once $G(T)/G_0$ exceeds the separating value of $1/2$ and $g_1(T)/g_0$ drops below $1/2$, see \fig{sym}. Within the Kondo regime ($0<J_Y<J_Y^*$) we can, thus, characterize the flow by a single Kondo scale $T_{\rm K}(J_Y)$ with $G_0[T_{\rm K}(J_Y)]=1/2$.
For non-zero RKKY-like coupling $J_Y\neq 0$, $T_{\rm K}(J_Y)$ is suppressed below its single-impurity value $T_{\rm K}(J_Y=0)=T_{\rm K}^0$ and ceases to exist beyond the critical
coupling $J_Y^*$. We find $T_{\rm K}(J_Y^*)=T_{\rm K}^0/e$, where
$e=2.718\dots$ is Euler's constant, in agreement with the analytic result of
Ref.~\cite{Hans} (see Fig.~3 of Ref.~\cite{KWJK_2imp}). Thus, $T_{\rm K}(J_Y)$
is a crossover scale which can be identified with a renormalized,
single-impurity Kondo temperature which remains finite at the transition.
By contrast, the strong-coupling scale $T_{{\rm Ks}}$ drops to $0$ 
quadratically at the Kondo-to-RKKY QPT as a function of $J_Y$, 
in agreement with the Kondo destruction hypothesis 
\cite{Si2001Oct,Coleman2001Aug,Senthil2004Jan,Friedemann2009Jul},
here confirmed for the $2$-impurity case. 
Similarly, $T_Y(J_Y)$ vanishes quadratically at the QSL transition,
see insets of Fig.~3 (c) of Ref.~\cite{KWJK_2imp}.
When channel asymmetry is included, a significant channel
dependence of all the scales leads to qualitatively different regimes at $T>0$
which we analyze in \Sec{SSS}.

\subsection{Fixed point spectra and Fermi-liquid character}
\label{sec:FPspectra}

\begin{figure}
\includegraphics[width=\linewidth]{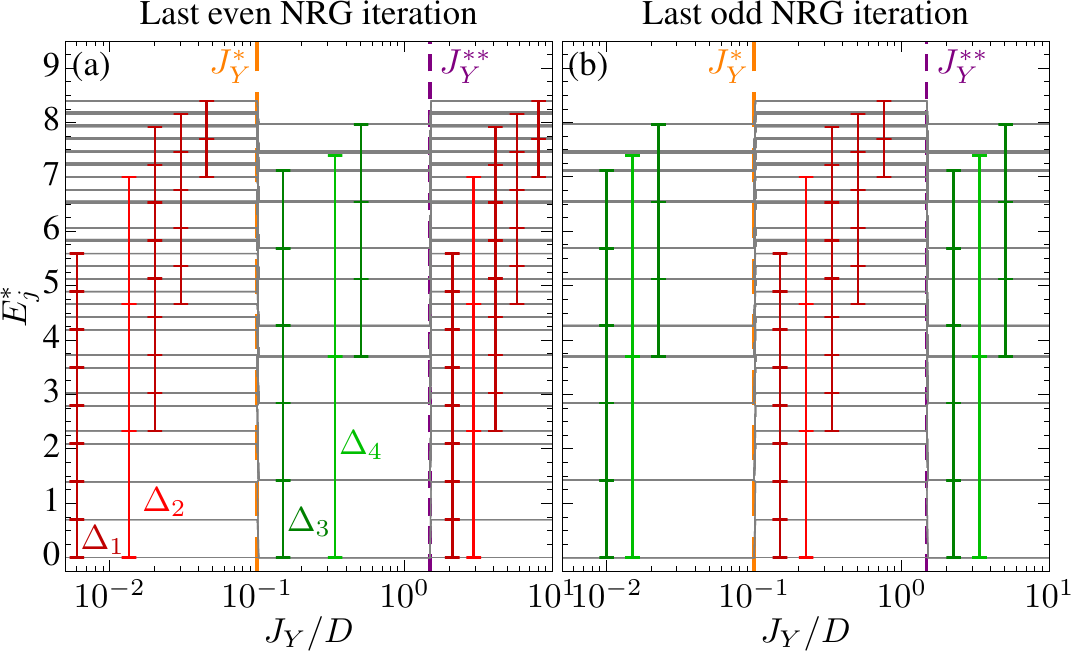}
\caption{Fixed point many-body spectra (1000 multiplets with lowest energy) 
		 at (a) last even and (b) last odd NRG iteration, for parameters 
		 as in \fig{sym} with $T=10^{-7}T_{\rm K}^0=10^{-11}U$, as functions of $J_Y$.  		 
		 The QPTs are indicated with dashed vertical lines.
		 All the many-body excitation energies can be composed of only $4$
		 single-particle excitation energies $\Delta_1$--$\Delta_4$ 
		 (see text for details).}
\label{fig:symE}
\end{figure}

Further insight into the structure of the Kondo, RKKY, and QSL phases can be gained 
from the fixed-point spectra in theses phases, \ie{}, the spectra of the (scaled) 
renormalized Hamiltonian to which the system converges for a large number of NRG 
iterations. As is well known, these spectra differ for even or odd number of NRG 
iterations, because in each iteration one site is added to a
Wilson chain and the total electron number (per channel) increases by one, 
so that the many-body ground state is alternatingly a spin singlet or a Kramers 
doublet \cite{WilsonNRG,NRG_RMP,Eickhoff2020Nov}.  

The fixed-point spectra of the symmetric model, \eq{Hsym}, are displayed
in \fig{symE} as function of $J_Y$ for the last even and last odd iteration,
clearly showing the two phase transitions as discontinuities
at $J_Y=J_Y^*$ and $J_Y^{**}$, respectively.
We see that all three stable phases are of Fermi liquid (FL) nature,
(but not the critical fixed points marking the transitions). 
This can be recognized from the fact, that the eigenenergies $E_j^*$
of the truncated fixed-point Hamiltonian up to about $j_{\rm max} = 1000$ multiplets
can be constructed as a linear combination of only two level spacings,
namely, for even iterations
$E_j^*=n\Delta_1+m\Delta_2$ in the Kondo phase ($J_Y<J_Y^*$) and
$E_j^*=n\Delta_3+m\Delta_4$ in the RKKY phase ($J_Y^*<J_Y<J_Y^{**}$) with
integer $n$, $m$. This is indicated in \fig{symE} (a) by the vertical
rulers. It means that the many-body states with energies $E_j^*$ consist
of multiple independent excitations of the same energies $\Delta_i$, the
quasiparticle excitations characteristic of a FL.
For the chosen $\Lambda=2.5$, we have $\Delta_1=0.699$, $\Delta_2=2.332$ and
$\Delta_3=1.422$, $\Delta_4=3.698$.

In fact, in the Kondo phase
the fixed-point spectrum is identical to that of the free conduction system
($\varGamma=0$, $J_Y=0$, not shown) which is a FL by definition.
It results from the fact that effectively one conduction electron from each
channel is used to form the Kondo singlet with the respective impurity
and the remaining conduction electrons are free particles with an impurity
scattering phase shift of $\varphi=\pi/2$ at the fixed point.
In the RKKY phase, the fixed-point spectrum is the same as in the case of 
free particles. 
Here, the Kondo screening is absent, thus, the impurity scattering phase 
shift $\varphi=0$. 
However, because of the Heisenberg interaction $J_Y$ within the 
conduction electron system, the relevant quasiparticles are superpositions
of free Bloch electrons, still forming FL excitations.
Finally, it is interesting to see in
\fig{symE} that the fixed-point spectra of the Kondo and the RKKY phases
are interchanged when one considers the odd- instead of even-iteration
spectra. This simply stems from the fact that in the Kondo phase 
one conduction electron is effectively removed from the system by the
Kondo screening as compared to the RKKY phase,
and the same is true when one considers the system
in the $(N-1)$st (odd) iteration instead of the $N$th (even) iteration
(see above).

\begin{figure}[t!]
\includegraphics[width=\linewidth]{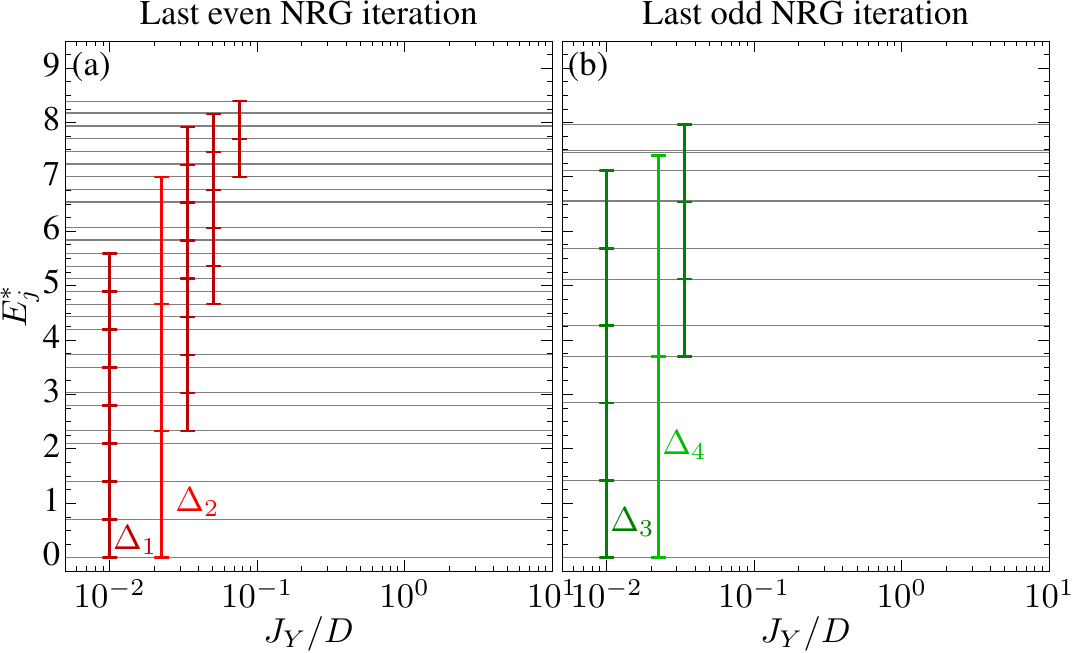}
\caption{Like \fig{symE}, except for $\varGamma=0.103U$ corresponding to $T_{\rm K}^0\approx 10^{-2}U$, 
		 and with $T=10^{-7}T_{\rm K}^0=10^{-9}U$. 
		 There are no QPTs in this case. 
		 Excitation energies $\Delta_1$--$\Delta_4$ are the same as in \fig{symE}.
		 %Parameters: $\Lambda=2.5$, $N=3000$, $6.5 < E_{\rm cut} < 7.0$.
		 }
\label{fig:symE_cross}
\end{figure}

Coming now to the QSL phase ($J_Y>J_Y^{**}$), it is striking in \fig{symE}
that the QSL fixed-point spectrum is identical to the Kondo fixed-point
spectrum. This means, in particular, that the QSL is also a FL. However,
it should be stressed at this point that the identical fixed point 
spectrum does not mean that the two phases are the same. The 
eigenstates are different compositions of the original degrees of freedom, 
in particular leading to a spreading of the phase shift into the 
interacting part of the conduction band and consequently non-universal 
values of impurity spectral density. 
This can be interpreted as non-universal fractionalization of the 
FL quasiparticles into conduction-band and impurity parts.
The former are responsible for the partial Kondo-screening of the impurities.
The latter give rise to  
interimpurity spin correlations of the QSL (see also \Sec{sym0}).
Nevertheless, the identical spectra indicate the possibility 
that the two phases can be continuously connected by circumventing the $T=0$
phase transitions in parameter space.
This scenario is indeed realized for strong enough
Kondo coupling $\varGamma$, measured by $T_{\rm K}^0$, as is shown by the fixed-point spectrum in \fig{symE_cross} for $T_{\rm K}^0\approx 10^{-2}\,U$
and by the general phase phase diagram in Fig.~2 (b) of Ref.~\cite{KWJK_2imp}.
The Kondo-QSL crossover corresponds to a continuous change of the proportions 
between the two parts of the quasi-particles driven by increasing $J_Y$.
In this case the Kondo fraction of them starts melting around $J_Y\sim D$
and the inter-channel fraction dominates completely only in the 
$J_Y\to\infty$ limit.

\subsection{Significance of the interaction in the band}
\label{sec:sym0}

\begin{figure}[t!]
\includegraphics[width=0.85\linewidth]{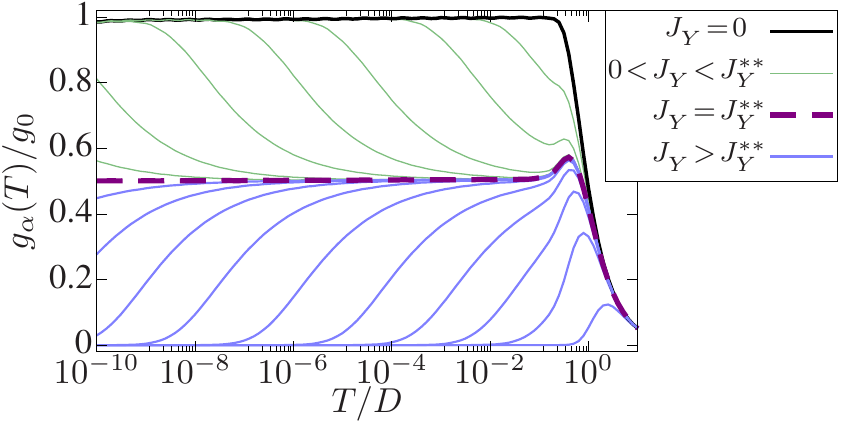}
\caption{Conduction band conductance $g_\alpha$ in a special 
		 case of $\varGamma_1 = \varGamma_2 = 0$, as a function of 
		 temperature, for different $J_Y$ in the range indicated 
		 in the legend. 
		 Other parameters as in \fig{sym}.
		 }
\label{fig:noQDs-flow}
\end{figure}

In conventional impurity models, the case of vanishing Kondo 
coupling is trivially a Fermi liquid (plus the decoupled impurities).
However in the case of the Hamiltonian given by \eq{Hsym},
even for $\varGamma_1=\varGamma_2=\varGamma=0$ there is an interaction term
proportional to $J_Y$, acting within the conduction band.
Therefore, let us consider now the case $\varGamma=0$.
After NRG mapping onto the Wilson chains, this model is very similar
to the Jones-Varma model \cite{Jones2} except that, instead of the
localized spins, $J_Y$ couples two sites strongly hybridizing with
the bath and possessing charge as well 
as spin degrees of freedom, since the Coulomb repulsion is absent. 
Despite these differences, an analogue of the Jones-Varma 
transition is still present 
and actually corresponds to the spin-liquid transition at 
$J_Y=J_Y^{**} \approx 1.5D$. 
It is driven by the spin exchange destabilizing the Bloch 
electrons when overcoming the corresponding bandwidth.
This is illustrated in \fig{noQDs-flow},
showing the conductance $g_\alpha$ as a function of $T$.
However, in the absence of impurities, the non-universal 
character of the phase for $J_Y>J_Y^{**}$ is lost,
for the impurity fraction of the quasiparticles cannot exist then
(see previous section). Instead,
a soft-gapped FL emerges, fully characterized by a density of states
featuring universal $\Acb(\w)\sim \w^2$ behavior.

The lack of conduction band electrons near the Fermi level
comes from the fact that this state may also be seen as an analogue
of a Kondo state, where each of the two conduction bands 
plays the role of a spin-screening channel for the other. This binds the 
electrons from the Fermi level into spin singlets,
as it does in the Kondo effect. The important difference is that, 
unlike the Kondo coupling, $J_Y$ does not flow under RG transformations,
thus, a critical value of $J_Y$ must be exceeded for this phase 
to form.
On the other hand, it should be noted that in the absence of
interactions within the conduction band, coupling the impurity to a
host with a static, not dynamically generated density of states
proportional to $\w^2$, does not lead to the screening of the impurity 
spin, even in the presence of strong Kondo coupling 
\cite{dosNRG,Fritz2004Dec,AndersNatPhys}. Hence, it is the competition
between both dynamical effects, the tendency to form the
interchannel screening induced by $J_Y$ and the Kondo effect induced by
non-zero $\varGamma$, that generates the frustration
stabilizing the QSL phase in the presence of Kondo impurities.
In this Kondo-stabilized QSL the impurity spectral density acquires
a nonuniversal value at $\w=0$. The screening of each impurity by the 
respective conduction band and the interimpurity local 
spin compensation both contribute to the non-universal 
ground state correlations. 
The scattering phase shift of the free part of conduction 
band electrons gets spread between the impurities and the
interacting part of the conduction band. For sufficiently 
strong $\varGamma$, decreasing $J_Y$ then leads to a 
continuous crossover to the Kondo state, with each impurity 
screened by the respective conduction band.  
The above observation points to the expectation
that, in order to capture a possible QSL of the type discussed here
in Kondo lattice systems, it is essential to treat both effects dynamically,
the Kondo screening and the spin correlations within the conduction band.
The latter is not performed in single-site DMFT \cite{DMFT_RMP}.

%%%%%%%%%%%%%%%%%%%%%%%%%%%%%%%%%%%%%%%%%%%%%%%%%%%%%%%%%%%%%%%%%%%%%%%%%%%%%%%%
\section{Asymmetry}
\label{sec:results}
%%%%%%%%%%%%%%%%%%%%%%%%%%%%%%%%%%%%%%%%%%%%%%%%%%%%%%%%%%%%%%%%%%%%%%%%%%%%%%%%

In the present section we analyze each type of asymmetry,
case by case, showing that the general scenario obtained for the symmetric
case survives at $T=0$ as long as there is no charge transfer.
Moreover, even in the presence of the latter many features of the symmetric
model can be identified in extended regimes, corresponding to the sharp phases 
of the symmetric model. However, channel asymmetry splits the Kondo 
scale into two different values for each channel, $T_{\rm K}^0 \to 
T_{{\rm K}1}^0,\,T_{{\rm K}2}^0$, and a large difference between the two will create
space for qualitatively new behavior at $T>0$.

%%%%%%%%%%%%%%%%%%%%%%%%%%%%%%%%%%%%%%%%%%%%%%%%%%%%%%%%%%%%%%%%%%%%%%%%%%%%%%%%
\subsection{Results for channel asymmetry}
\label{sec:SSS}
%%%%%%%%%%%%%%%%%%%%%%%%%%%%%%%%%%%%%%%%%%%%%%%%%%%%%%%%%%%%%%%%%%%%%%%%%%%%%%%%

\begin{figure}[tb]
\includegraphics[width=\linewidth]{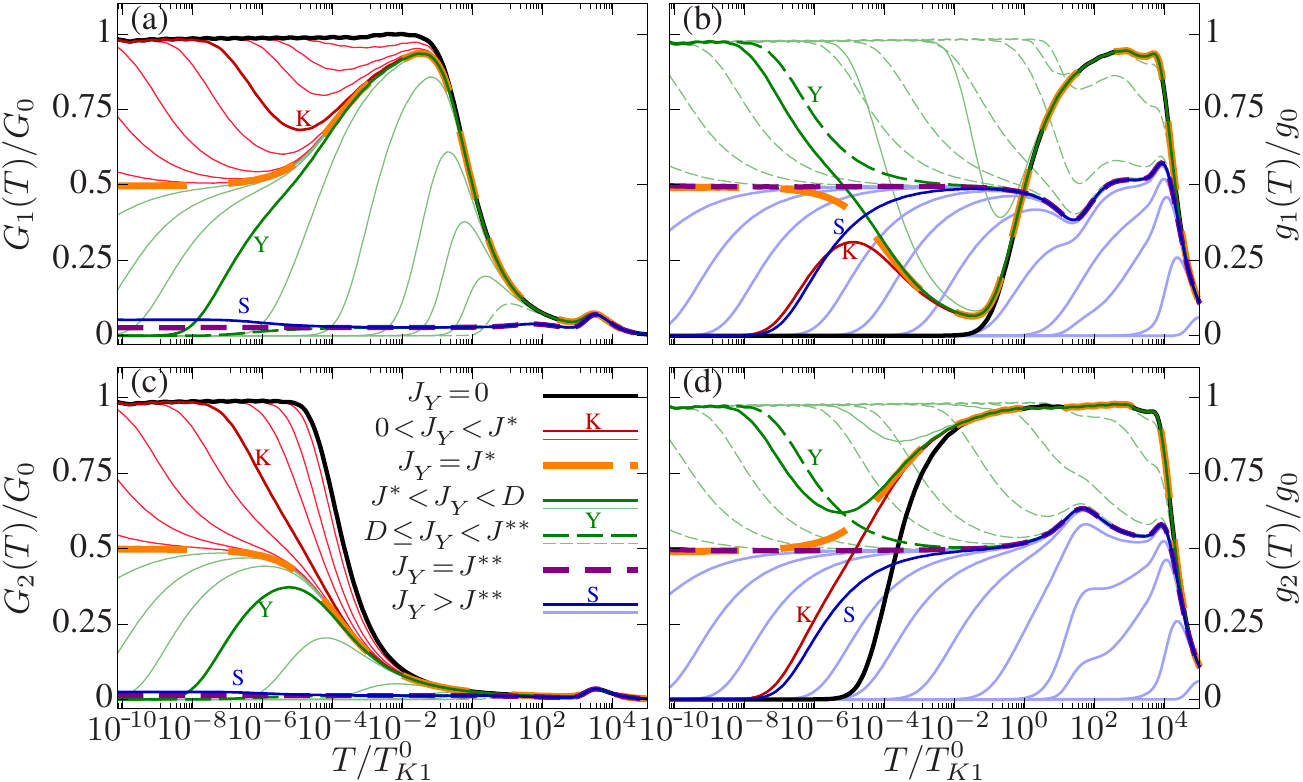}
\caption{The flow diagrams of 
		 the impurity conductances (a) $G_1$ and (c) $G_2$
                 as well as the conducton-band conductances
                 (b) $g_1$, and (d) $g_2$,
		 in presence of channel asymmetry induced by a
		 difference in the Kondo couplings, $\varGamma_2 = 0.5\,\varGamma_1 = 0.0244 U$.
		 Other parameters: $D=2U$, $\Lambda=2.5$, $6.5 < E_{\rm cut} <
                 7$. Representative flow curves towards the Kondo, RKKY, and
                 SQL fixed points are marked with K, Y, and S, respectively.}
\label{fig:asym12-flow}
\end{figure}

\subsubsection{General picture}

As the first type of asymmetry we study the channel asymmetry. 
Let us start by inspecting the flow diagram for 
$\varGamma_2 = \varGamma_1/2$, which is presented in \fig{asym12-flow}. 
First of all, we clearly see that reducing $\varGamma_2$ by a factor of $2$
drives the corresponding Kondo temperature $T_{{\rm K}2}^0$ down by almost 
$4$ orders of magnitude, a manifestation of its exponential
dependence on the coupling strength, cf.~\fig{asym12-flow}~(c).
This separation of energy scales facilitates analyzing the regime 
$T_{{\rm K}2} \ll T \ll T_{{\rm K}1}$ by NRG, see below. 
For smaller asymmetry, this intermediate regime is realized at higher temperatures.
For  $T < \min(T_{{\rm K}1},T_{{\rm K}2})$
all the properties of the symmetric model \cite{KWJK_2imp} are recovered.
In particular, the impurity conductances exhibit universal values: 
$G_1(T\!=\!0) = G_2(T\!=\!0) = G_0$ in the Kondo regime ($J_Y<J_Y^*$)  
and $G_1(0) = G_2(0) = 0$ in the RKKY phase ($J_Y^* < J_Y^{} < J_Y^{**}$). 
The two phases are separated by a Jones-Varma QPT \cite{Jones2},
\ie{}, the unstable QPT fixed point at $J_Y^{} = J_Y^*$
where $G_1(0) = G_2(0) = G_0/2$. 
Further increase of $J_Y$ drives the spin-liquid QPT \cite{KWJK_2imp}
at $J_Y =J_Y^{**}\approx 1.5D$,
where the values of $G_1(0)$ and $G_2(0)$ become non-universal. 

For a Fermi liquid in the presence of PH symmetry, the only two possible values 
of scattering phase shift from an impurity are $0$ or $\pi/2$ 
\cite{phi_book,Affleck}. However, the $J_Y$ term in \eq{Hsym} acting 
between the hosts does not correspond to a single-particle dispersion,
but introduces interactions into the conduction band. 
For this reason, $G_\alpha$, $\alpha=1,\,2$, does not have to 
assume a universal value at $T\to 0$. 
Nevertheless, as this is the only interacting 
term in the leads, the remaining parts of the electrodes 
(\ie{} each entire electrode $\alpha$ excluding the state on which  
$J_Y$ directly acts) are still a FL, and
the phase-shift argument is valid for $g_\alpha(0)$.
Therefore, $g_\alpha(0)$ always assumes universal values $0$ or $g_0$,
except for the critical points at $J_Y^{} = J_Y^*$ and $J_Y^{} = J_Y^{{**}}$,
where non-Fermi-liquid critical fixed points allow for 
$g_1(0) = g_2(0) = g_0/2$. Thus, together with $G_\alpha(0)$, 
the values of $g_\alpha(0)$ characterize all the phases uniquely, in
exactly the same way as for the channel-symmetric case \cite{KWJK_2imp}.

\begin{figure}[t!]
\includegraphics[width=\linewidth]{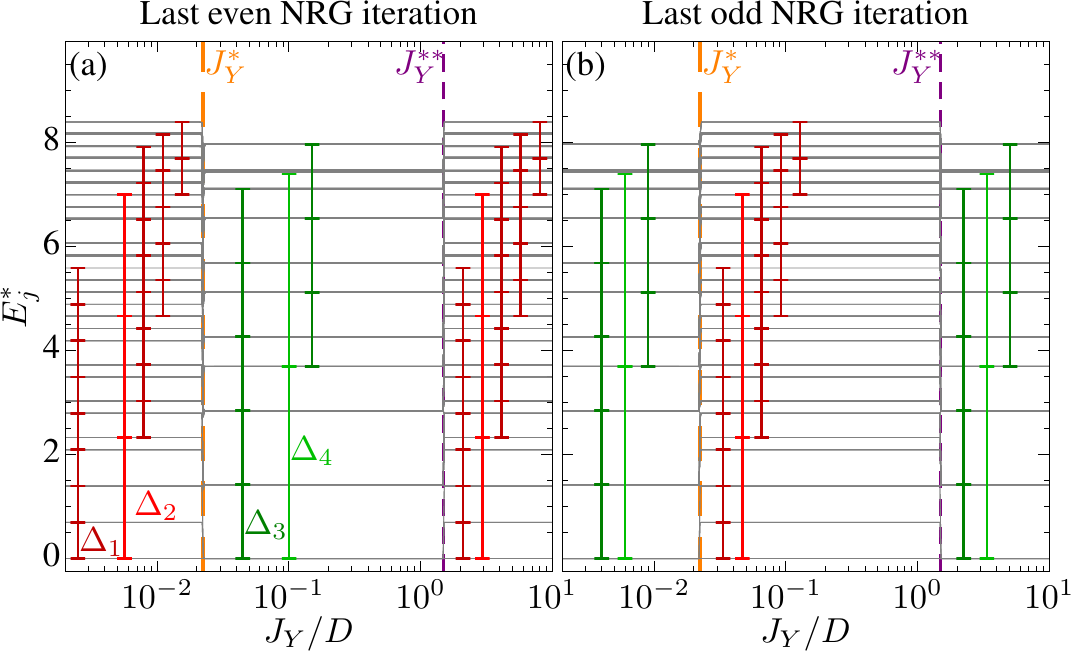}
\caption{Fixed-point spectra at the (a) last even and (b) last odd NRG iteration
for parameters as in \fig{asym12-flow} with $T=10^{-7}T_{\rm K}^0=10^{-11}U$
as functions of $J_Y$. The QPT points are indicated by vertical, dashed lines.
}
\label{fig:asym12-E}
\end{figure}

\begin{figure}[b!]
\includegraphics[width=\linewidth]{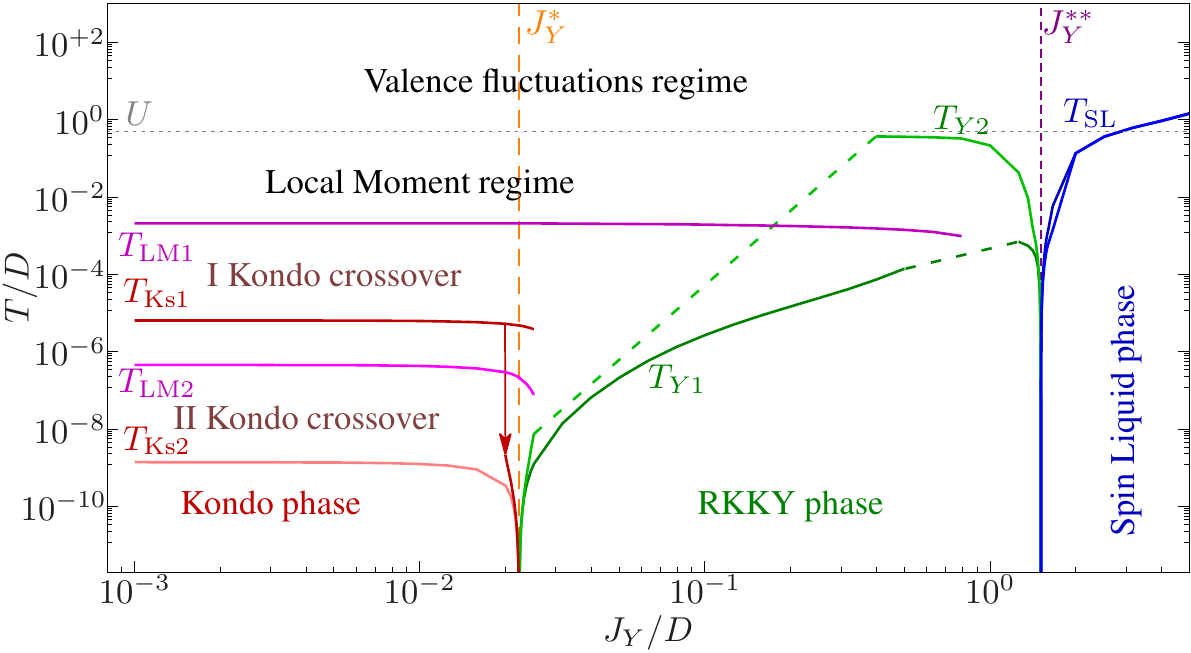}
\caption{Phase diagram obtained from the NRG flow of \fig{asym12-flow}. 
The ambiguity of $T_{{\rm Ks}1}$ (indicated by the vertical arrow next to the 
$J_Y^*$ line) is caused by the fact 
that the Kondo regime is approached twice.  $T_{Y1}$ and $T_{Y2}$ are not 
well defined far from the transitions and represented as dashed, straight
lines. In the QSL phase, $T_{{\rm SL}1}=T_{{\rm SL}2}=T_{\rm SL}$. }
\label{fig:asym12-scales}
\end{figure}

While at $T=0$ the regimes of the channel-asymmetric model 
are the same as in the symmetric one, the situation changes 
at $T>0$. For $T_{{\rm K}2} < T < T_{{\rm K}1}$,
the second impurity is still in the local-moment regime, such
that neither the Kondo effect nor the RKKY interaction 
are significant there. Meanwhile, the first impurity 
is already screened by its host. In such conditions $T_{{\rm K}1}^0$
is almost not affected by $J_Y$, as seen in \fig{asym12-flow}~(a). 
Only at lower $T \approx T_{{\rm K}2}^0$, $G_1(T)$ drops from a 
value close to $G_0$, characteristic of the Kondo regime,
towards the unstable QPT fixed-point value $G_0/2$. Upon further 
decreasing $T\to 0$, in the Kondo phase ($J_Y<J_Y^*$) $G_1(T)$ returns 
to the unitary value $G_1(0)=G_0$ and in the RKKY phase 
($J_Y^{*}<J_Y^{}<J_Y^{**}$) it further drops to $G_1(0)=0$. In this way,
the stronger-coupled impurity approaches the Kondo regime
twice: once at $T\lesssim T_{{\rm K}1}$, and then again
at $T < T_{{\rm Ks}2}$.
In between, at $T\sim T_{{\rm K}2}$ the second impurity leaves 
the local-moment regime, and its flow toward the Kondo regime affects 
the screening in the first channel via the RKKY interaction,
which becomes relevant in this temperature range. Hence, for the 
first, stronger-coupled impurity, $T_{{\rm Ks}1}$ is not unique then, 
and only one of these two strong-coupling Kondo scales vanishes at 
the Jones-Varma transition as in the symmetric case, while the other 
one does not, cf. \fig{asym12-flow}~(a).
On the other hand, for the second impurity, $T_{{\rm Ks}2}$ is more efficiently 
suppressed to zero than in the symmetric case, cf. \fig{asym12-flow}~(c) .
The host conductances $g_\alpha$ behave as can be expected on the 
basis of the symmetric case. They reveal all the characteristic 
energy scales in their temperature dependencies. As can be seen
in \figs{asym12-flow}, $g_\alpha(T)/g_0 \approx 1-G_\alpha(T) / G_0$
except for the QSL phase, where $g_\alpha$ vanishes
despite a small, non-universal $G_\alpha$. The fixed-point spectra 
for the parameter values as in \fig{asym12-flow} are shown  
in \fig{asym12-E}, confirming that the Fermi liquid nature of the 
three different phases is preserved in the channel-asymmetric case, 
see the discussion in \Sec{FPspectra}.

Summarizing the asymmetric NRG flow of \fig{asym12-flow}, the phase diagram in 
the $J_Y$--$T$ plane is presented in \fig{asym12-scales}. 
Note that the scales $T_{{\rm LM}\alpha}$, $\alpha=1,\,2$, persist across the 
Jones-Varma transition and the strong-coupling scale of the weaker-coupled 
impurity, $T_{{\rm K}s2}$ vanishes, as expected from the symmetric case. However,
the low-energy scale of the 
stronger-coupled impurity, $T_{{\rm K}s1}$, splits into two values, 
$T_{{\rm K}s2}'>T_{{\rm K}s2}$ where the larger one remains finite across the QPT. 
This means that the expected scaling behavior near the QPT will 
exist only for $T<T_{{\rm K}}'$, which should be experimentally observable 
in asymmetric setups.
By contrast, both $T_{Y\alpha}$ scales vanish at the QSL transition.

\subsubsection{Dependence on asymmetries 
$\varGamma_2/\varGamma_1$ and $D_2/D_1$}

\begin{figure}[t!]
\includegraphics[width=\linewidth]{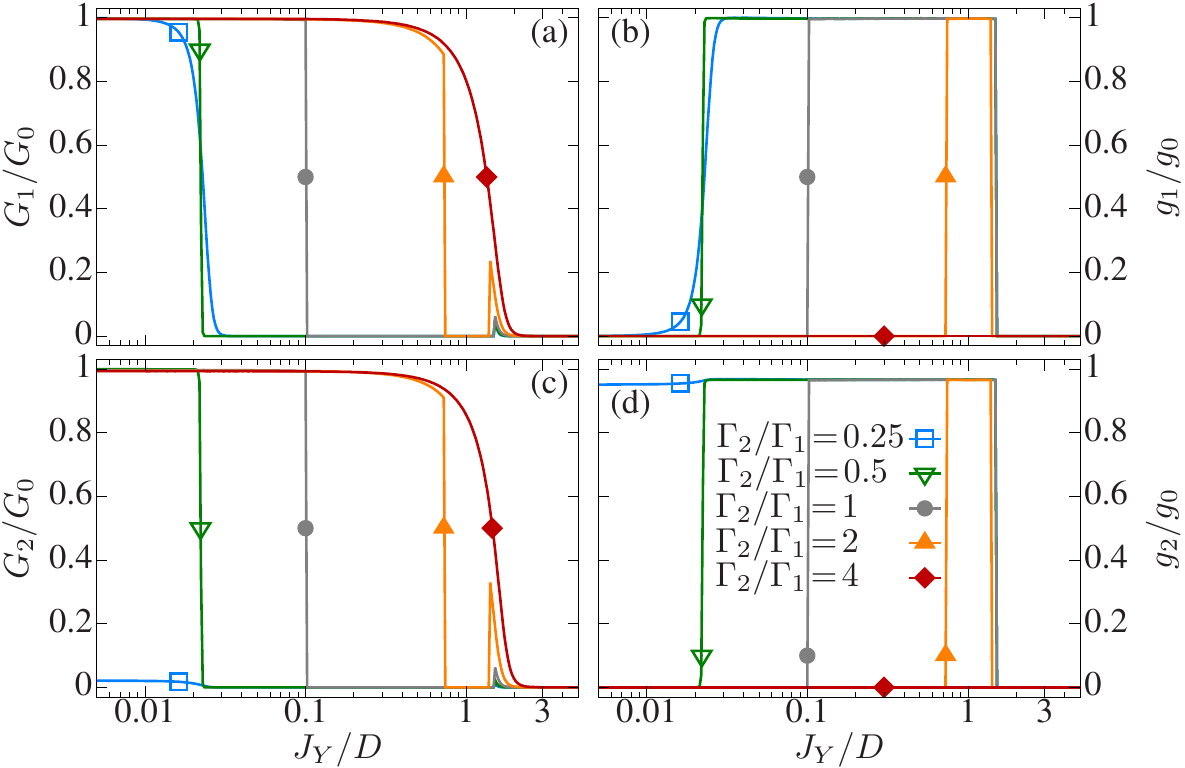}
\caption{The conductances $G_\alpha$ (left panels) and $g_\alpha$ (right
  panels), $\alpha=1,\,2$, as functions of $J_Y$ for $T=10^{-11}U$ and various
  ratios of the Kondo couplings $\varGamma_2/\varGamma_1$ for fixed 
$\varGamma_1=0.0488\,U$. Other parameters are as in \fig{asym12-flow}.
}
\label{fig:asym12-1D}
\end{figure}

Let us now inspect $G_\alpha$ and $g_\alpha$ as functions of $J_Y$ at 
a cryogenic but non-zero temperature $T=10^{-11}U$, see \fig{asym12-1D}.
For $\varGamma_2 = 2\,\varGamma_1$
there are two phase transitions similar to the symmetric 
case ($\varGamma_2=\varGamma_1$), as already discussed in the context 
of \figs{asym12-flow} and \ref{fig:asym12-scales}. With decreasing 
$\varGamma_2/\varGamma_1$ at fixed $\varGamma_1=0.0488\,U$ the Jones-Varma 
phase boundary $J_Y^*$ shifts towards smaller values. We can estimate
this decrease analytically in the following way. 
In the presence of asymmetric Kondo couplings there exist two different   
single-impurity Kondo scales, see also \fig{asym12-flow}. 
For a PH-symmetric Anderson impurity model 
($\varepsilon_{\alpha}=U/2$), they are approximately given by \cite{Haldane}
\beq
T_{{\rm K}\alpha}^0=\sqrt{U\varGamma_{\alpha}}\,e^{-\pi U/8\varGamma_\alpha}, \qquad
\alpha=1,\,2. 
\label{eq:TKalpha}
\eeq
The RKKY coupling can be calculated perturbatively as
\beq
Y\simeq (\rho_1 J_{{\rm K}1}) (\rho_2 J_{{\rm K}2}) J_Y,
\label{eq:RKKY}
\eeq
with the Kondo spin-exchange coupling $\rho_{\alpha}J_{{\rm K}\alpha}=
4\rho_{\alpha} V_{\alpha}^2/U=4\varGamma_{\alpha}/\pi U$.
Generalizing the Doniach criterion for 
RKKY-induced Kondo breakdown \cite{Doniach} to channel asymmetry, 
the state with two individually Kondo screened impurities is expected 
to terminate when $Y$ reaches a critical strength which is roughly 
given by the smaller one of the two Kondo scales.
This is confirmed by the
NRG flow in \fig{asym12-flow}, where it is the smaller one of the two scales
that determines which fixed point is reached at the 
lowest energy, with some modification of $T_{{\rm K}\alpha}$ by the 
inter-channel coupling $J_Y$. Combining this criterion with 
\eqs{eq:TKalpha}{eq:RKKY} we obtain the critical inter-channel coupling
for the asymmetric Jones-Varma transition ($\varGamma_2<\varGamma_1$),      
\beq
J_Y^* \simeq \frac{(\pi U)^2}{4\varGamma_1}\, \sqrt{\frac{U}{\Gamma_2}}\,
e^{-\pi U/8\varGamma_2},
\eeq
which is exponentially suppressed with $\varGamma_2/U$. When $T_{{\rm K}2}$ 
becomes smaller than the temperature $T$ of the system, the transition 
gets smeared as seen in \fig{asym12-1D} for $\varGamma_2/\varGamma_1= 0.25$.

Note that the QSL transition persists even for the smallest 
$\varGamma_2/\varGamma$ values, since its characteristic energy scale is
fixed at $J_Y^{**}\approx 1.5\,D\gg T$ \cite{KWJK_2imp}. 
On the other hand, when the asymmetry $\varGamma_2/\varGamma_1$ 
exceeds a certain threshold, $J_Y^*$ exceeds $J_Y^{**}$, \ie{}, the 
intermediate RKKY phase vanishes, and the two QPTs merge to a single
Kondo-to-QSL crossover \cite{KWJK_2imp}, as seen in \fig{asym12-1D} for 
$\varGamma/\varGamma_1=4$.

To summarize this section, the $\varGamma_2 \neq \varGamma_1$ case 
is qualitatively the same as the symmetric case at $T=0$, as long 
as the effective scale for Kondo breakdown does not exceed the 
threshold value for changing the $2$-QPTs scenario to the no-QPT scenario. 
However, at elevated $T$ a novel 
regime appears, when one of the impurities is practically decoupled.
The RKKY interaction and the Kondo effect on the more weakly coupled 
impurity are not relevant there, while the QSL phase remains
robust, unless one of the impurities is completely detached. 
The extreme asymmetric limit of decoupling one impurity is subtle and
requires further study beyond the scope of the present article.
We further find that asymmetry in the bandwidths, $D_2/D_1\neq 1$, 
(not shown) causes effects similar to the coupling asymmetry. 
It leads to different Kondo scales $T_{{\rm K}\alpha}$ and crossover 
scales $T_{Y\alpha}$ \new{etc.}, but does not drive qualitatively new phenomena.

\begin{figure}[b!]
\includegraphics[width=\linewidth]{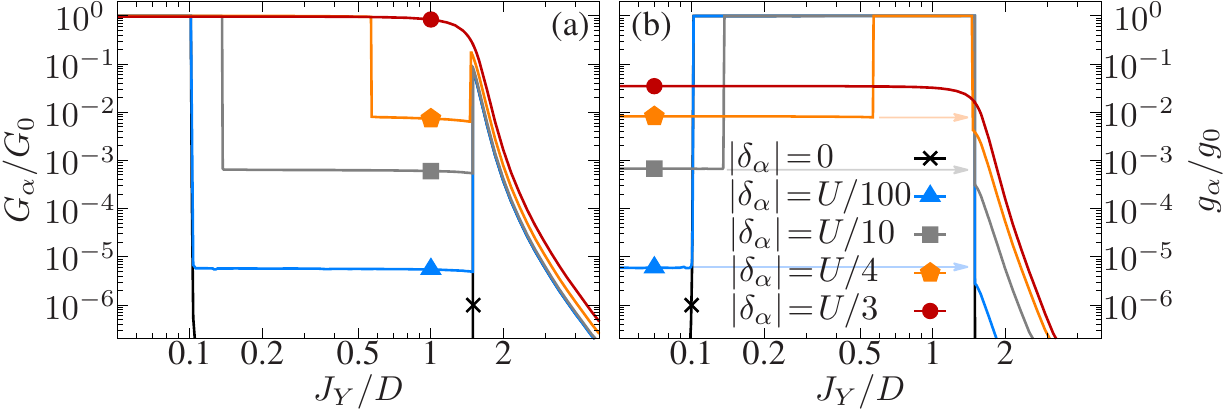}
\caption{$G_1=G_2$ and $g_1=g_2$ as functions of $J_Y$ for 
$T_{\rm K}^0=10^{-7}\,U$, $T=10^{-11}\,U$, $\Lambda = 2.5$, and 
PH asymmetry $|\delta_1|=|\delta_2|$. Other parameters are as in \fig{sym}. 
  The results do not depend on the sign of $\delta_{\alpha}$.}
\label{fig:asymEps-1D}
\end{figure}

%%%%%%%%%%%%%%%%%%%%%%%%%%%%%%%%%%%%%%%%%%%%%%%%%%%%%%%%%%%%%%%%%%%%%%%%%%%%%%%%
\subsection{Results for particle-hole asymmetry}
\label{sec:SUU}
%%%%%%%%%%%%%%%%%%%%%%%%%%%%%%%%%%%%%%%%%%%%%%%%%%%%%%%%%%%%%%%%%%%%%%%%%%%%%%%%

We now focus on the PH asymmetry as defined in \Sec{modelSUU},
still in the absence of inter-channel charge transfer. We consider first the
channel-symmetric case, $\delta_1=\delta_2=\delta$, 
so that $G_1=G_2$ and analogous for all other physical quantities. 

\fig{asymEps-1D} showns that the two QPTs, visible as discontinuities 
of the conductances $G_1$ and $g_1$ as functions of $J_Y$ near zero 
temperature, $T=11^{-11}\,U$, are robust against PH symmetry breaking, 
just as the Jones-Varma transition is in the case of 
direct inter-impurity spin exchange, see Ref.~\cite{Zarand2006Oct}.
With increasing PH asymmetry $\delta$ the first, 
Jones-Varma-like QPT shifts from $J_Y^*\approx 0.1\,D$ to larger critical 
values $J_Y^*$, while the second QPT at $J_Y^{**}\approx 1.5\,D$ is 
independent of $\delta$, until both QPTs merge into a single 
crossover for $\delta\gtrsim U/3$ [red curves in \fig{asymEps-1D} 
(a), (b)]. This behavior can be understood in the following way. 
According to the Doniach criterion \cite{Doniach} the Kondo breakdown
occurs when the effective RKKY interaction $Y$ exceeds the single-impurity
Kondo scale $T_{\rm K}^{0}$, where $Y\sim [\rho J_{{\rm K}}(\delta)]^2J_Y$ and
$J_{{\rm K}}(\delta)=(\varGamma / \pi) [U/(U^2/4 - \delta^2)]$ from
a Schrieffer-Wolff transformation \cite{Haldane,Hewson_book} of the Hamiltonian 
\eq{Hsym}. PH-asymmetry thus leads to a squared exponential  
increase of the Kondo scale,
\beq
T_{{\rm K}}^0(\delta) =  
		T_{{\rm K}}^0(0)\, e^{\pi\delta^2 / (2\varGamma U) },
\label{TKvsDelta}
\eeq
and, therefore, to an increase of the critical $J_Y^*(\delta)$ of the
Jones-Varma transition, independent of the sign of $\delta$. 
The merging of the two QPTs to a single crossover with increasing 
$T_{\rm  K}(0)$, i.e., the vanishing of the RKKY phase at a critical 
point $T_{\rm K\,max}^0$ was explained for the PH-symmetric case in \Sec{sym} and 
observed in the phase diagram of Ref.~\cite{KWJK_2imp}, Fig.~2. 
Here, it is an experimentally relevant observation that one can switch 
between the single-crossover and the double-QPT scenarios by tuning the
PH asymmetry, \eg, by gating the impurity levels.

From \fig{asymEps-1D} we see that $G_{\alpha}/G_0\approx 1$ and 
$g_\alpha \approx 0$ for $J_Y<J_Y^*$. 
However, with $|\delta_\alpha| > 0$ small non-universality appears, 
characteristic of asymmetric Anderson model \cite{KMWWb},
well visible as small $g_{\alpha}>0$ values, even in the $T\to 0$ limit.
Similarly, in the 
RKKY phase ($J_Y^*<J_Y<J_Y^{**}$)
$G_{\alpha}$ takes nonzero values. These are signatures of nonuniversal 
scattering phase shifts of the band electrons in both regimes.
The persistence of true QPTs even in this 
nonuniversal situation is a special feature of the two-host model \eq{Hsym},
designed to describe an effective RKKY interaction without charge transfer. 
It is in stark contrast to the conventional single-host $2$-impurity model, 
where the QPT can be restored by a fine-tuned counter-term only if the 
phase shifts from the two impurities compensate each other \cite{Fabian}. 

\begin{figure}[t!]
\includegraphics[width=\linewidth]{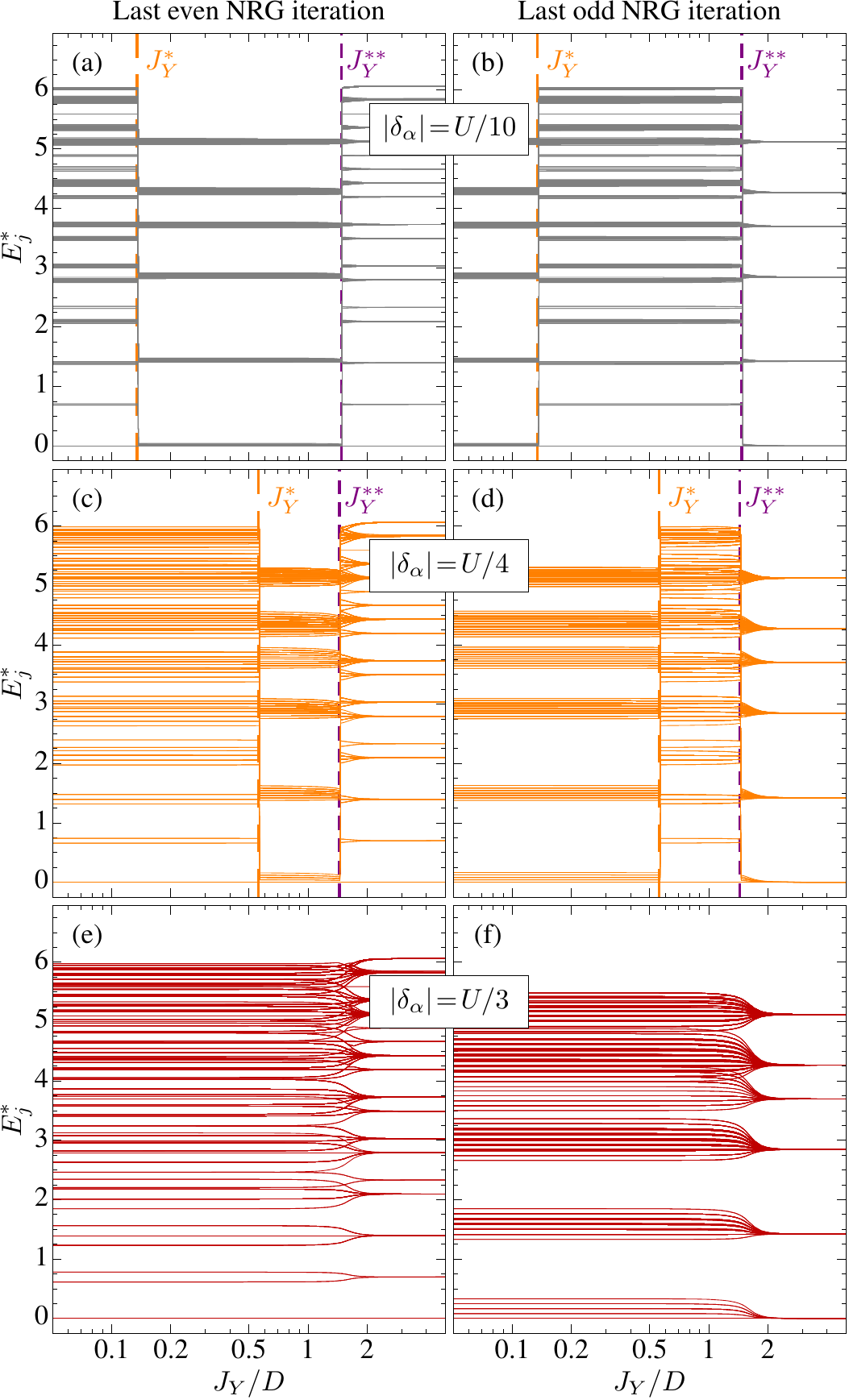}
\caption{Fixed point spectra (lowest $1000$ states) for PH asymmetry, 
$\delta_{\alpha} \neq 0$. Parameters and a color code are the 
same as in \fig{asymEps-1D}, with 
(a-b) $|\delta_{\alpha}|=U/10$, 
(c-d) $|\delta_{\alpha}|=U/4$,
(e-f) $|\delta_{\alpha}|=U/3$. 
Note that the $\delta_{\alpha}=0$ case is presented in \fig{symE}.
}
\label{fig:asymEps-E}
\end{figure}

The above properties are corroborated by the fixed-point spectra shown 
in \fig{asymEps-E}. The persisting QPTs are clearly visible as 
discontinuities of the spectra below the crossover threshold, 
$T_{\rm K}^{0}<T_{\rm K\,max}^{0}$, \fig{asymEps-E} (a)--(d). However, the
spectra are no longer universal as in the fully symmetric case. Instead,
the low-lying multiplets get split, a manifestation that each
$\delta_1$, $\delta_2$ renders a marginally relevant perturbation to 
the PH-symmetric fixed point, extending the latter into a surface 
of fixed points, analogous to the line of fixed points in 
the case of a single impurity Anderson model \cite{KMWWb}.
Above the threshold, $T_{\rm K}^{0}>T_{\rm K\,max}^{0}$, the spectra 
show a crossover from the Kondo to the QSL regime, as expected  
[\fig{asymEps-E} (e), (f)].

\begin{figure}[t!]
\includegraphics[width=\linewidth]{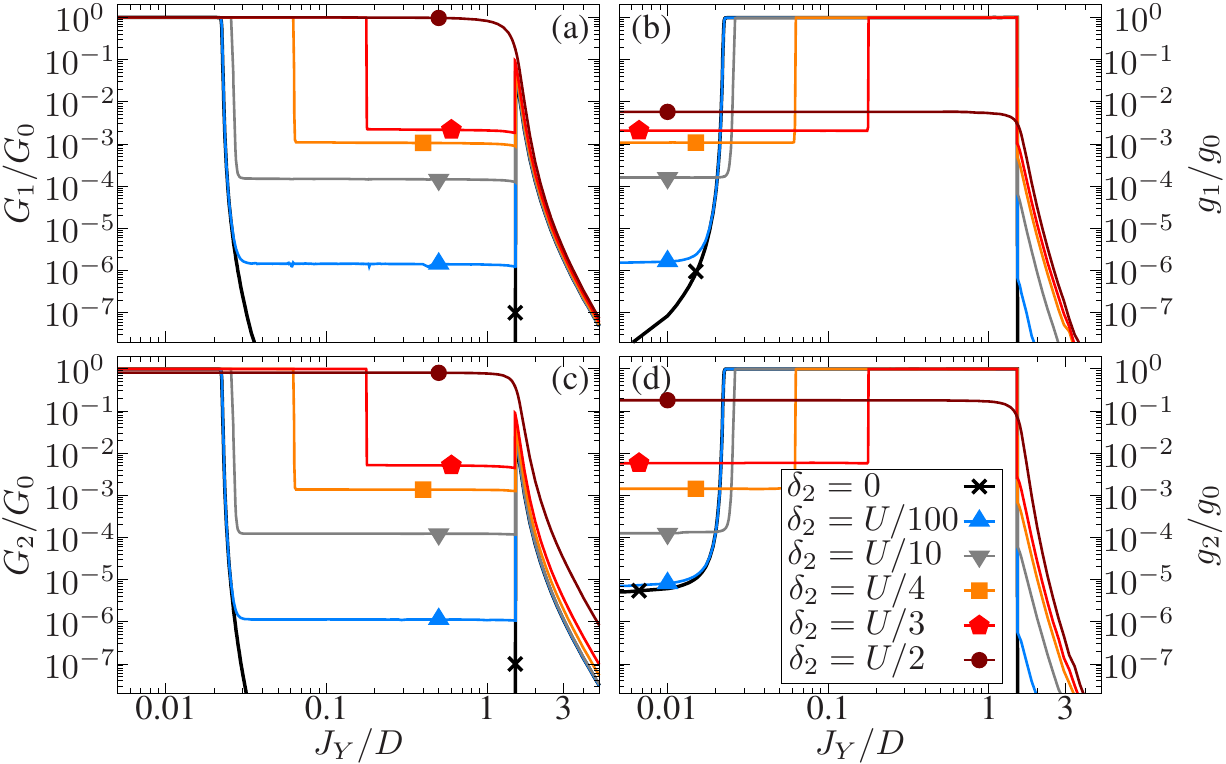}
\caption{Conductances as functions of $J_Y$ for $\varGamma_2 = \varGamma_1 =
  0.0488U$ and $\delta_2=2 \delta_1$. Other parameters as in \fig{asymEps-1D}.
		 }
\label{fig:asymEps12-1D}
\end{figure}

Finally, in \fig{asymEps12-1D} we show the results for the channel-asymmetric
case in addition to PH asymmetry, $\delta_\alpha \neq 0$. 
As seen in the figure, the effects of both types of asymmetry just ``add up''.
PH asymmetry induces non-universality, while channel asymmetry
is relevant mainly at elevated temperatures and does not lead to qualitatively 
new features. Imposing $|\delta_1| \neq |\delta_2|$ does not induce 
qualitative changes, albeit the level of deviation of $G_\alpha$ from 
universality is substantially increased in the channel with 
larger $\delta_\alpha$.

%%%%%%%%%%%%%%%%%%%%%%%%%%%%%%%%%%%%%%%%%%%%%%%%%%%%%%%%%%%%%%%%%%%%%%%%%%%%%%%%
\subsection{Results in the presence of charge transfer}
\label{sec:SS}
%%%%%%%%%%%%%%%%%%%%%%%%%%%%%%%%%%%%%%%%%%%%%%%%%%%%%%%%%%%%%%%%%%%%%%%%%%%%%%%%

\begin{figure}[b!]
\includegraphics[width=\linewidth]{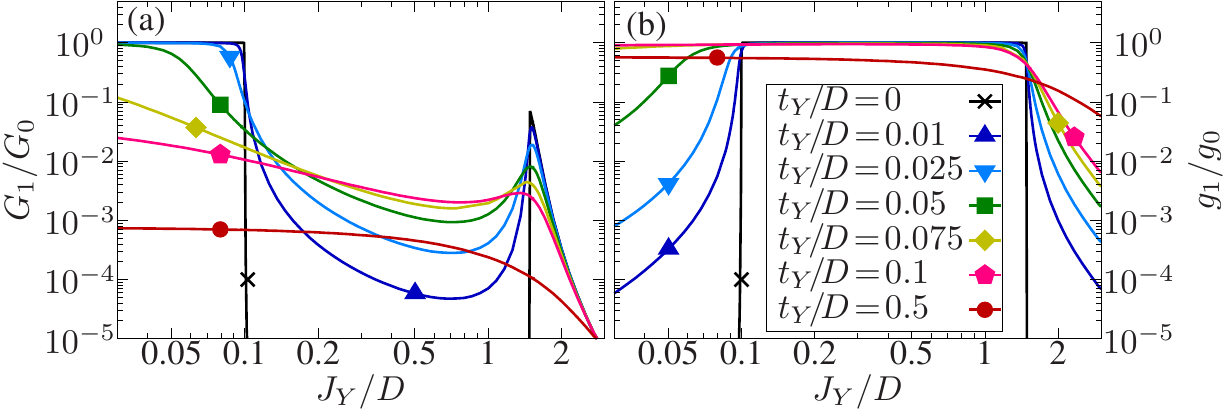}
\caption{$T=0$ conductances $G_1(0)$ and $g_1(0)$ as functions of $J_Y$ for
  different inter-site hoppings $t_Y$, and $U_{\rm cb} = U/100$. 
Other parameters as in \fig{asymEps-1D} with $\delta_\alpha=0$,
except $\Lambda=3$ and $5.5<E_{\rm cut}<6$.}
\label{fig:asym-1D}
\end{figure}

In this section we address the case with charge transfer 
allowed between the channels by $t_Y\neq 0$, as introduced in \Sec{modelSSZ}, 
and all other parameters symmetric. 
The most important results concerning charge transfer are presented 
in \fig{asym-1D}, where $G_1=G_2$ and $g_1=g_2$ are plotted as functions of 
$J_Y$. The NRG parameters used there, $\Lambda=3$ and $5.5<E_{\rm cut}<6$,
are slightly modified in comparison to earlier figures, to reduce the 
numerical effort increased by lifting the intra-host $U(1)$ symmetries. 

\begin{figure}[t!]
\includegraphics[width=\linewidth]{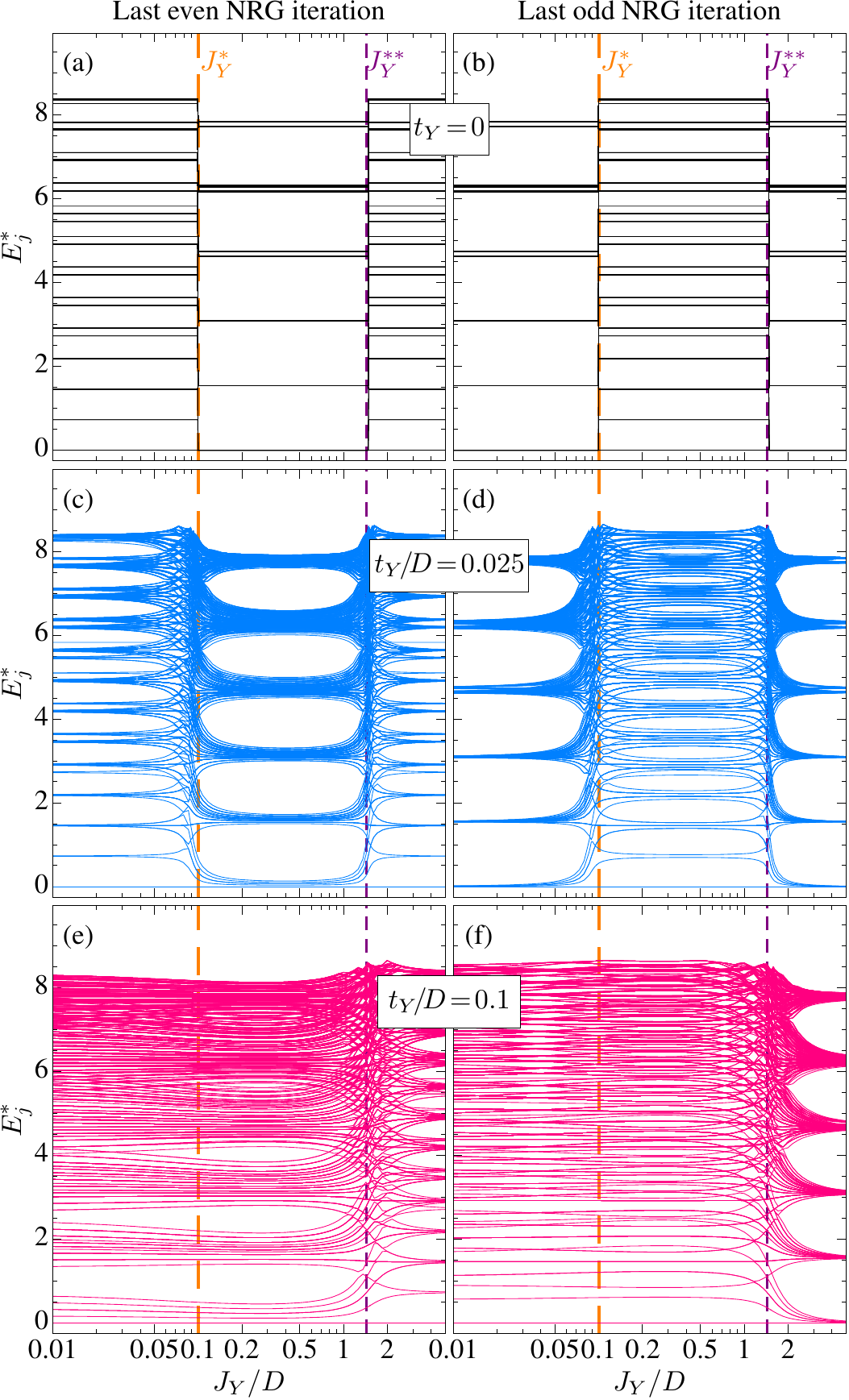}
\caption{The fixed point spectra (lowest $1000$ states) for (a-b) $t_Y=0$,
		 (c-d) $t_Y=0.025D$ and $t_Y/D=0.1$ (e-f). 
		 Vertical lines indicate the positions of QPTs at $t_Y=0$.
		 Parameters as in \fig{asym-1D}. %, with $U_{\rm cb}=U/100$.
		 Note that due to $\Lambda=3$ the energy levels cannot be 
		 directly compared to Figs.~\ref{fig:symE}, \ref{fig:asymEps-E} etc.}
\label{fig:asym-E}
\end{figure}

Without charge transfer, the Jones-Varma and the QSL QPTs exist
at $J_Y=J_Y^*\approx 0.1\,D$, in agreement with the Doniach critical value of
the RKKY coupling $Y^* \simeq [4\Gamma/(\pi\rho U)]^2J_Y^* \approx T_K^0$,
and at $J_Y=J_Y^{**}\approx 1.5\,D$, corresponding to the
characteristic quasiparticle energies in the Kondo and the RKKY phase,
respectively \cite{KWJK_2imp}. 
For $t_Y\neq 0$, both QPTs get smeared into a crossover \cite{Fye,Affleck},
although $t_Y$ also generates a contribution to the RKKY coupling.
One can see from the logarithmic $J_Y$ scale in \fig{asym-1D} that
for both crossovers and fixed $t_Y$,
the crossover width scales roughly with the respective
critical value, $J_Y^*$ or $J_Y^{**}$. 
Noteworthy, $t_Y$ destroys the universality features in both, the Kondo and 
the RKKY regimes by introducing a marginally relevant perturbation 
around the respective fixed points. Consequently, $G_1$ ($g_1$) does 
not reach unity in the Kondo (RKKY) regime, and does not drop to 0 
in the RKKY (Kondo) regime. This loss of universality is further 
illustrated by the fixed point spectra in \fig{asym-E}, where the
levels, which are degenerate for $t_Y=0$ (a-b), are split
for $t_Y\neq 0$ (c-f), in addition to the smearing of the QPTs.

The smearing of the QPTs raises the question about their 
experimental observability in the presence of charge transfer. 
For the QSL transition, the inter-host magnetic coupling must reach
$J_Y\gtrsim 1.5\,D$. This seems possible 
in low-bandwidth systems, especially in magic-angle, twisted bilayer 
graphene \cite{TBG1,TBG2}, doped with Kondo impurities and with an
additional magnetic exchange $J_Y$ between the hosts.
However, magnetic exchange is usually accompanied by charge transfer.
In \fig{asym-1D} the crossovers are still well pronounced up
to $t_Y/J_{Yc}\approx 0.1$, where $J_{Yc}=J_Y^*\approx 0.1\, D$ for the
Jones-Varma and $J_{Yc}=J_Y^{**}\approx 1.5\, D$ for the QSL transition,
but are washed out for larger $t_Y/J_{Yc}$ values. This indicates that
the QSL transition/crossover allows for significantly larger absolute
values of charge transfer to be observed than the Jones-Varma transition.
The narrow width of the QSL crossover for up to
$t_Y\lesssim 0.05\,D$ (see \fig{asym-1D})
gives rise to the expectation that this crossover should be well
observable in two-impurity tunneling setups \cite{Bork} 
or by reducing charge transfer by superexchange coupling of the hosts.

More important, however, would be the existence of a Kondo stabilized
QSL in HF lattice systems. There, the metallic hosts of our model
are represented by a single band with equal hopping matrix elements $t$
between all neighboring sites. Charge transfer between lattice sites is,
therefore, an integral part of the band. A recent 
DMFT study of the Anderson lattice with antiferromagnetic order
competing with the Kondo effect \cite{Gleis} mapped such
a lattice on a two-impurity model, where DMFT self-consistency 
restored the Jones-Varma-like QPT, even though 
charge transfer was present in the effective two-impurity model. 
This is similar to the restoration of the QPT by a fine-tuned
counter-term as in Ref.~\cite{Fabian}, with the difference that the
DMFT self-consistency (antiferromagnetic order) tunes the system to cancel
the marginally relevant charge transfer operator, which otherwise would
destabilize the transition.
However, the study of Ref.~\cite{Fabian} did not include genuine
magnetic interactions within the conduction electron system as represented
by the Heisenberg term $J_Y$ in our model, \eq{Hsym}, which is instrumental
in stabilizing the QSL phase. We, therefore, expect that in HF materials
a QSL phase may be stabilized by strong,
antiferromagnetic correlations within the conduction band (distinct
from the Kondo-induced correlations), as they are
generated, \eg, near a spin-density-wave instability.

%%%%%%%%%%%%%%%%%%%%%%%%%%%%%%%%%%%%%%%%
%%%%%%%%%%%%%%%%%%%%%%%%%%%%%%%%%%%%%%%%
\section{Conclusion}
\label{sec:con}
%%%%%%%%%%%%%%%%%%%%%%%%%%%%%%%%%%%%%%%%
%%%%%%%%%%%%%%%%%%%%%%%%%%%%%%%%%%%%%%%%

We have analyzed the influence of various types of asymmetry on a
generic $2$-impurity, $2$-host Anderson system in the Kondo limit. 
We found that, as long as there is no charge transfer between the two 
screening channels, the results 
of the symmetric case apply qualitatively at sufficiently low temperatures. 
In particular, generically the system exhibits   
two QPTs. The Jones-Varma QPT occurs at a critical
value $J_Y^*$ when the 
inter-host spin exchange $J_Y$ drives the RKKY coupling $Y$ 
to overcome the effective Kondo scale of the system $T_{\rm K}(J_Y)$, 
destroying Kondo quasi-particles. 
The second QPT is found for $J_Y$ exceeding the critical 
value $J_Y^{**}$, close to the conduction bandwidth $D$. In the 
case of sufficiently strong Kondo couplings, $J_Y^*$ may exceed $J_Y^{**}$,
so that both QPTs merge to a single crossover. 

If the Kondo couplings are not equal, two separate Kondo scales,
$T_{{\rm K}1}^0$, $T_{{\rm K}2}^0$, appear in the NRG flow
(each corresponding to one channel), and the position of the Jones-Varma
QPT is roughly determined by the smaller one of the two
Kondo scales.
A novel regime appears in an elevated temperature range defined by
$T_{{\rm K}2} < T < T_{{\rm K}1}$ when the Kondo scales of the two impurities
are sufficiently different. 
In this interesting regime, at $J_Y^{} \ll J_Y^{**}$, the system behaves as 
if the weaker-coupled impurity were detached. 
However, the QSL phase transition is robust, unless one of the impurities 
gets completely detached. 
Even PH asymmetry by independently gating the impurity levels does not
lead to smearing of the QPTs. Instead, only the spectral densities
acquire nonuniversal values. 
A similar behavior is expected for PH-asymmetric hosts.
In line with Ref.~\cite{Zarand2006Oct}, this gives hope for the experimental 
realization of the transitions in quantum-dot nanostructures. 

We expect that a quantum spin liquid of the metallic type discussed here
may be stabilized in heavy-fermion systems, near a spin-density wave
instability of the conduction band, where strong antiferromagnetic
correlations within the conduction band compete with Kondo screening
of the localized moments, analogous to the inter-host spin coupling
in the present two-impurity model.

\begin{acknowledgments}
Stimulating discussions with Frithjof Anders, Fabian Eickhoff, Andreas Gleis,
Mohsen Hafez-Torbati, and Kacper Wrze\'{s}niewski are gratefully acknowledged.
This project was financially supported by the Deutsche Forschungsgemeinschaft
(DFG, German Research Foundation) under Germany's Excellence Strategy –
Cluster of Excellence \textit{Matter and Light for Quantum Computing}, ML4Q
(390534769), and through the DFG Collaborative Research Center CRC 185 OSCAR
(277625399). K.P.W. acknowledges funding by the Alexander von Humboldt
Foundation and support from the Polish National Science Centre through grant no.~2018/29/B/ST3/00937. 
\end{acknowledgments}


\begin{thebibliography}{66}%
\makeatletter
\providecommand \@ifxundefined [1]{%
 \@ifx{#1\undefined}
}%
\providecommand \@ifnum [1]{%
 \ifnum #1\expandafter \@firstoftwo
 \else \expandafter \@secondoftwo
 \fi
}%
\providecommand \@ifx [1]{%
 \ifx #1\expandafter \@firstoftwo
 \else \expandafter \@secondoftwo
 \fi
}%
\providecommand \natexlab [1]{#1}%
\providecommand \enquote  [1]{``#1''}%
\providecommand \bibnamefont  [1]{#1}%
\providecommand \bibfnamefont [1]{#1}%
\providecommand \citenamefont [1]{#1}%
\providecommand \href@noop [0]{\@secondoftwo}%
\providecommand \href [0]{\begingroup \@sanitize@url \@href}%
\providecommand \@href[1]{\@@startlink{#1}\@@href}%
\providecommand \@@href[1]{\endgroup#1\@@endlink}%
\providecommand \@sanitize@url [0]{\catcode `\\12\catcode `\$12\catcode
  `\&12\catcode `\#12\catcode `\^12\catcode `\_12\catcode `\%12\relax}%
\providecommand \@@startlink[1]{}%
\providecommand \@@endlink[0]{}%
\providecommand \url  [0]{\begingroup\@sanitize@url \@url }%
\providecommand \@url [1]{\endgroup\@href {#1}{\urlprefix }}%
\providecommand \urlprefix  [0]{URL }%
\providecommand \Eprint [0]{\href }%
\providecommand \doibase [0]{https://doi.org/}%
\providecommand \selectlanguage [0]{\@gobble}%
\providecommand \bibinfo  [0]{\@secondoftwo}%
\providecommand \bibfield  [0]{\@secondoftwo}%
\providecommand \translation [1]{[#1]}%
\providecommand \BibitemOpen [0]{}%
\providecommand \bibitemStop [0]{}%
\providecommand \bibitemNoStop [0]{.\EOS\space}%
\providecommand \EOS [0]{\spacefactor3000\relax}%
\providecommand \BibitemShut  [1]{\csname bibitem#1\endcsname}%
\let\auto@bib@innerbib\@empty
%</preamble>
\bibitem [{\citenamefont {Coleman}\ and\ \citenamefont
  {Nevidomskyy}(2010)}]{Coleman_FrustrationAndKondoInHF}%
  \BibitemOpen
  \bibfield  {author} {\bibinfo {author} {\bibfnamefont {P.}~\bibnamefont
  {Coleman}}\ and\ \bibinfo {author} {\bibfnamefont {A.~H.}\ \bibnamefont
  {Nevidomskyy}},\ }\bibfield  {title} {\bibinfo {title} {{Frustration and the
  Kondo Effect in Heavy Fermion Materials}},\ }\href
  {https://doi.org/10.1007/s10909-010-0213-4} {\bibfield  {journal} {\bibinfo
  {journal} {J. Low Temp. Phys.}\ }\textbf {\bibinfo {volume} {161}},\ \bibinfo
  {pages} {182} (\bibinfo {year} {2010})}\BibitemShut {NoStop}%
\bibitem [{\citenamefont {Kirchner}\ \emph {et~al.}(2020)\citenamefont
  {Kirchner}, \citenamefont {Paschen}, \citenamefont {Chen}, \citenamefont
  {Wirth}, \citenamefont {Feng}, \citenamefont {Thompson},\ and\ \citenamefont
  {Si}}]{Kirchner2020Mar}%
  \BibitemOpen
  \bibfield  {author} {\bibinfo {author} {\bibfnamefont {S.}~\bibnamefont
  {Kirchner}}, \bibinfo {author} {\bibfnamefont {S.}~\bibnamefont {Paschen}},
  \bibinfo {author} {\bibfnamefont {Q.}~\bibnamefont {Chen}}, \bibinfo {author}
  {\bibfnamefont {S.}~\bibnamefont {Wirth}}, \bibinfo {author} {\bibfnamefont
  {D.}~\bibnamefont {Feng}}, \bibinfo {author} {\bibfnamefont {J.~D.}\
  \bibnamefont {Thompson}},\ and\ \bibinfo {author} {\bibfnamefont
  {Q.}~\bibnamefont {Si}},\ }\bibfield  {title} {\bibinfo {title} {{Colloquium:
  Heavy-electron quantum criticality and single-particle spectroscopy}},\
  }\href {https://doi.org/10.1103/RevModPhys.92.011002} {\bibfield  {journal}
  {\bibinfo  {journal} {Rev. Mod. Phys.}\ }\textbf {\bibinfo {volume} {92}},\
  \bibinfo {pages} {011002} (\bibinfo {year} {2020})}\BibitemShut {NoStop}%
\bibitem [{\citenamefont {Doniach}(1977)}]{Doniach}%
  \BibitemOpen
  \bibfield  {author} {\bibinfo {author} {\bibfnamefont {S.}~\bibnamefont
  {Doniach}},\ }\bibfield  {title} {\bibinfo {title} {{The Kondo lattice and
  weak antiferromagnetism}},\ }\href
  {https://doi.org/10.1016/0378-4363(77)90190-5} {\bibfield  {journal}
  {\bibinfo  {journal} {Physica B+C}\ }\textbf {\bibinfo {volume} {91}},\
  \bibinfo {pages} {231} (\bibinfo {year} {1977})}\BibitemShut {NoStop}%
\bibitem [{\citenamefont {Ruderman}\ and\ \citenamefont {Kittel}(1954)}]{RK}%
  \BibitemOpen
  \bibfield  {author} {\bibinfo {author} {\bibfnamefont {M.~A.}\ \bibnamefont
  {Ruderman}}\ and\ \bibinfo {author} {\bibfnamefont {C.}~\bibnamefont
  {Kittel}},\ }\bibfield  {title} {\bibinfo {title} {{Indirect Exchange
  Coupling of Nuclear Magnetic Moments by Conduction Electrons}},\ }\href
  {https://doi.org/10.1103/PhysRev.96.99} {\bibfield  {journal} {\bibinfo
  {journal} {Phys. Rev.}\ }\textbf {\bibinfo {volume} {96}},\ \bibinfo {pages}
  {99} (\bibinfo {year} {1954})}\BibitemShut {NoStop}%
\bibitem [{\citenamefont {Kasuya}(1956)}]{K}%
  \BibitemOpen
  \bibfield  {author} {\bibinfo {author} {\bibfnamefont {T.}~\bibnamefont
  {Kasuya}},\ }\bibfield  {title} {\bibinfo {title} {{A Theory of Metallic
  Ferro- and Antiferromagnetism on Zener's Model}},\ }\href
  {https://doi.org/10.1143/PTP.16.45} {\bibfield  {journal} {\bibinfo
  {journal} {Prog. Theor. Phys.}\ }\textbf {\bibinfo {volume} {16}},\ \bibinfo
  {pages} {45} (\bibinfo {year} {1956})}\BibitemShut {NoStop}%
\bibitem [{\citenamefont {Yosida}(1957)}]{Y}%
  \BibitemOpen
  \bibfield  {author} {\bibinfo {author} {\bibfnamefont {K.}~\bibnamefont
  {Yosida}},\ }\bibfield  {title} {\bibinfo {title} {{Magnetic Properties of
  Cu-Mn Alloys}},\ }\href {https://doi.org/10.1103/PhysRev.106.893} {\bibfield
  {journal} {\bibinfo  {journal} {Phys. Rev.}\ }\textbf {\bibinfo {volume}
  {106}},\ \bibinfo {pages} {893} (\bibinfo {year} {1957})}\BibitemShut
  {NoStop}%
\bibitem [{\citenamefont {Jones}\ and\ \citenamefont {Varma}(1987)}]{Jones1}%
  \BibitemOpen
  \bibfield  {author} {\bibinfo {author} {\bibfnamefont {B.~A.}\ \bibnamefont
  {Jones}}\ and\ \bibinfo {author} {\bibfnamefont {C.~M.}\ \bibnamefont
  {Varma}},\ }\bibfield  {title} {\bibinfo {title} {{Study of two magnetic
  impurities in a Fermi gas}},\ }\href
  {https://doi.org/10.1103/PhysRevLett.58.843} {\bibfield  {journal} {\bibinfo
  {journal} {Phys. Rev. Lett.}\ }\textbf {\bibinfo {volume} {58}},\ \bibinfo
  {pages} {843} (\bibinfo {year} {1987})}\BibitemShut {NoStop}%
\bibitem [{\citenamefont {Jones}\ \emph {et~al.}(1988)\citenamefont {Jones},
  \citenamefont {Varma},\ and\ \citenamefont {Wilkins}}]{Jones2}%
  \BibitemOpen
  \bibfield  {author} {\bibinfo {author} {\bibfnamefont {B.~A.}\ \bibnamefont
  {Jones}}, \bibinfo {author} {\bibfnamefont {C.~M.}\ \bibnamefont {Varma}},\
  and\ \bibinfo {author} {\bibfnamefont {J.~W.}\ \bibnamefont {Wilkins}},\
  }\bibfield  {title} {\bibinfo {title} {{Low-Temperature Properties of the
  Two-Impurity Kondo Hamiltonian}},\ }\href
  {https://doi.org/10.1103/PhysRevLett.61.125} {\bibfield  {journal} {\bibinfo
  {journal} {Phys. Rev. Lett.}\ }\textbf {\bibinfo {volume} {61}},\ \bibinfo
  {pages} {125} (\bibinfo {year} {1988})}\BibitemShut {NoStop}%
\bibitem [{\citenamefont {Jones}\ and\ \citenamefont {Varma}(1989)}]{Jones3}%
  \BibitemOpen
  \bibfield  {author} {\bibinfo {author} {\bibfnamefont {B.~A.}\ \bibnamefont
  {Jones}}\ and\ \bibinfo {author} {\bibfnamefont {C.~M.}\ \bibnamefont
  {Varma}},\ }\bibfield  {title} {\bibinfo {title} {{Critical point in the
  solution of the two magnetic impurity problem}},\ }\href
  {https://doi.org/10.1103/PhysRevB.40.324} {\bibfield  {journal} {\bibinfo
  {journal} {Phys. Rev. B}\ }\textbf {\bibinfo {volume} {40}},\ \bibinfo
  {pages} {324} (\bibinfo {year} {1989})}\BibitemShut {NoStop}%
\bibitem [{\citenamefont {Gan}(1995)}]{Gan1995Mar}%
  \BibitemOpen
  \bibfield  {author} {\bibinfo {author} {\bibfnamefont {J.}~\bibnamefont
  {Gan}},\ }\bibfield  {title} {\bibinfo {title} {{Mapping the Critical Point
  of the Two-Impurity Kondo Model to a Two-Channel Problem}},\ }\href
  {https://doi.org/10.1103/PhysRevLett.74.2583} {\bibfield  {journal} {\bibinfo
   {journal} {Phys. Rev. Lett.}\ }\textbf {\bibinfo {volume} {74}},\ \bibinfo
  {pages} {2583} (\bibinfo {year} {1995})}\BibitemShut {NoStop}%
\bibitem [{\citenamefont {Mitchell}\ \emph {et~al.}(2012)\citenamefont
  {Mitchell}, \citenamefont {Sela},\ and\ \citenamefont
  {Logan}}]{Mitchell2012Feb}%
  \BibitemOpen
  \bibfield  {author} {\bibinfo {author} {\bibfnamefont {A.~K.}\ \bibnamefont
  {Mitchell}}, \bibinfo {author} {\bibfnamefont {E.}~\bibnamefont {Sela}},\
  and\ \bibinfo {author} {\bibfnamefont {D.~E.}\ \bibnamefont {Logan}},\
  }\bibfield  {title} {\bibinfo {title} {{Two-Channel Kondo Physics in
  Two-Impurity Kondo Models}},\ }\href
  {https://doi.org/10.1103/PhysRevLett.108.086405} {\bibfield  {journal}
  {\bibinfo  {journal} {Phys. Rev. Lett.}\ }\textbf {\bibinfo {volume} {108}},\
  \bibinfo {pages} {086405} (\bibinfo {year} {2012})}\BibitemShut {NoStop}%
\bibitem [{\citenamefont {Bork}\ \emph {et~al.}(2011)\citenamefont {Bork},
  \citenamefont {Zhang}, \citenamefont
  {Diekh{\ifmmode\ddot{o}\else\"{o}\fi}ner}, \citenamefont {Borda},
  \citenamefont {Simon}, \citenamefont {Kroha}, \citenamefont {Wahl},\ and\
  \citenamefont {Kern}}]{Bork}%
  \BibitemOpen
  \bibfield  {author} {\bibinfo {author} {\bibfnamefont {J.}~\bibnamefont
  {Bork}}, \bibinfo {author} {\bibfnamefont {Y.-h.}\ \bibnamefont {Zhang}},
  \bibinfo {author} {\bibfnamefont {L.}~\bibnamefont
  {Diekh{\ifmmode\ddot{o}\else\"{o}\fi}ner}}, \bibinfo {author} {\bibfnamefont
  {L.}~\bibnamefont {Borda}}, \bibinfo {author} {\bibfnamefont
  {P.}~\bibnamefont {Simon}}, \bibinfo {author} {\bibfnamefont
  {J.}~\bibnamefont {Kroha}}, \bibinfo {author} {\bibfnamefont
  {P.}~\bibnamefont {Wahl}},\ and\ \bibinfo {author} {\bibfnamefont
  {K.}~\bibnamefont {Kern}},\ }\bibfield  {title} {\bibinfo {title} {{A tunable
  two-impurity Kondo system in an atomic point contact}},\ }\href
  {https://doi.org/10.1038/nphys2076} {\bibfield  {journal} {\bibinfo
  {journal} {Nat. Phys.}\ }\textbf {\bibinfo {volume} {7}},\ \bibinfo {pages}
  {901} (\bibinfo {year} {2011})}\BibitemShut {NoStop}%
\bibitem [{\citenamefont {Bayat}\ \emph {et~al.}(2014)\citenamefont {Bayat},
  \citenamefont {Johannesson}, \citenamefont {Bose},\ and\ \citenamefont
  {Sodano}}]{Bayat2014May}%
  \BibitemOpen
  \bibfield  {author} {\bibinfo {author} {\bibfnamefont {A.}~\bibnamefont
  {Bayat}}, \bibinfo {author} {\bibfnamefont {H.}~\bibnamefont {Johannesson}},
  \bibinfo {author} {\bibfnamefont {S.}~\bibnamefont {Bose}},\ and\ \bibinfo
  {author} {\bibfnamefont {P.}~\bibnamefont {Sodano}},\ }\bibfield  {title}
  {\bibinfo {title} {{An order parameter for impurity systems at quantum
  criticality}},\ }\href {https://doi.org/10.1038/ncomms4784} {\bibfield
  {journal} {\bibinfo  {journal} {Nat. Commun.}\ }\textbf {\bibinfo {volume}
  {5}},\ \bibinfo {pages} {1} (\bibinfo {year} {2014})}\BibitemShut {NoStop}%
\bibitem [{\citenamefont {Pr{\ifmmode\ddot{u}\else\"{u}\fi}ser}\ \emph
  {et~al.}(2014)\citenamefont {Pr{\ifmmode\ddot{u}\else\"{u}\fi}ser},
  \citenamefont {Dargel}, \citenamefont {Bouhassoune}, \citenamefont {Ulbrich},
  \citenamefont {Pruschke}, \citenamefont {Lounis},\ and\ \citenamefont
  {Wenderoth}}]{Pruser2014Nov}%
  \BibitemOpen
  \bibfield  {author} {\bibinfo {author} {\bibfnamefont {H.}~\bibnamefont
  {Pr{\ifmmode\ddot{u}\else\"{u}\fi}ser}}, \bibinfo {author} {\bibfnamefont
  {P.~E.}\ \bibnamefont {Dargel}}, \bibinfo {author} {\bibfnamefont
  {M.}~\bibnamefont {Bouhassoune}}, \bibinfo {author} {\bibfnamefont {R.~G.}\
  \bibnamefont {Ulbrich}}, \bibinfo {author} {\bibfnamefont {T.}~\bibnamefont
  {Pruschke}}, \bibinfo {author} {\bibfnamefont {S.}~\bibnamefont {Lounis}},\
  and\ \bibinfo {author} {\bibfnamefont {M.}~\bibnamefont {Wenderoth}},\
  }\bibfield  {title} {\bibinfo {title} {{Interplay between the Kondo effect
  and the Ruderman{\textendash}Kittel{\textendash}Kasuya{\textendash}Yosida
  interaction}},\ }\href {https://doi.org/10.1038/ncomms6417} {\bibfield
  {journal} {\bibinfo  {journal} {Nat. Commun.}\ }\textbf {\bibinfo {volume}
  {5}},\ \bibinfo {pages} {1} (\bibinfo {year} {2014})}\BibitemShut {NoStop}%
\bibitem [{\citenamefont {Spinelli}\ \emph {et~al.}(2015)\citenamefont
  {Spinelli}, \citenamefont {Gerrits}, \citenamefont {Toskovic}, \citenamefont
  {Bryant}, \citenamefont {Ternes},\ and\ \citenamefont
  {Otte}}]{Spinelli2015Nov}%
  \BibitemOpen
  \bibfield  {author} {\bibinfo {author} {\bibfnamefont {A.}~\bibnamefont
  {Spinelli}}, \bibinfo {author} {\bibfnamefont {M.}~\bibnamefont {Gerrits}},
  \bibinfo {author} {\bibfnamefont {R.}~\bibnamefont {Toskovic}}, \bibinfo
  {author} {\bibfnamefont {B.}~\bibnamefont {Bryant}}, \bibinfo {author}
  {\bibfnamefont {M.}~\bibnamefont {Ternes}},\ and\ \bibinfo {author}
  {\bibfnamefont {A.~F.}\ \bibnamefont {Otte}},\ }\bibfield  {title} {\bibinfo
  {title} {{Exploring the phase diagram of the two-impurity Kondo problem}},\
  }\href {https://doi.org/10.1038/ncomms10046} {\bibfield  {journal} {\bibinfo
  {journal} {Nat. Commun.}\ }\textbf {\bibinfo {volume} {6}},\ \bibinfo {pages}
  {1} (\bibinfo {year} {2015})}\BibitemShut {NoStop}%
\bibitem [{\citenamefont {Moro-Lagares}\ \emph {et~al.}(2019)\citenamefont
  {Moro-Lagares}, \citenamefont {Koryt{\ifmmode\acute{a}\else\'{a}\fi}r},
  \citenamefont {Piantek}, \citenamefont {Robles}, \citenamefont {Lorente},
  \citenamefont {Pascual}, \citenamefont {Ibarra},\ and\ \citenamefont
  {Serrate}}]{Moro-Lagares2019May}%
  \BibitemOpen
  \bibfield  {author} {\bibinfo {author} {\bibfnamefont {M.}~\bibnamefont
  {Moro-Lagares}}, \bibinfo {author} {\bibfnamefont {R.}~\bibnamefont
  {Koryt{\ifmmode\acute{a}\else\'{a}\fi}r}}, \bibinfo {author} {\bibfnamefont
  {M.}~\bibnamefont {Piantek}}, \bibinfo {author} {\bibfnamefont
  {R.}~\bibnamefont {Robles}}, \bibinfo {author} {\bibfnamefont
  {N.}~\bibnamefont {Lorente}}, \bibinfo {author} {\bibfnamefont {J.~I.}\
  \bibnamefont {Pascual}}, \bibinfo {author} {\bibfnamefont {M.~R.}\
  \bibnamefont {Ibarra}},\ and\ \bibinfo {author} {\bibfnamefont
  {D.}~\bibnamefont {Serrate}},\ }\bibfield  {title} {\bibinfo {title} {{Real
  space manifestations of coherent screening in atomic scale Kondo lattices}},\
  }\href {https://doi.org/10.1038/s41467-019-10103-5} {\bibfield  {journal}
  {\bibinfo  {journal} {Nat. Commun.}\ }\textbf {\bibinfo {volume} {10}},\
  \bibinfo {pages} {1} (\bibinfo {year} {2019})}\BibitemShut {NoStop}%
\bibitem [{\citenamefont {Gleis}\ \emph {et~al.}(2022)\citenamefont {Gleis},
  \citenamefont {Lee}, \citenamefont {Weichselbaum}, \citenamefont {Kotliar},\
  and\ \citenamefont {von Delft}}]{Gleis}%
  \BibitemOpen
  \bibfield  {author} {\bibinfo {author} {\bibfnamefont {A.}~\bibnamefont
  {Gleis}}, \bibinfo {author} {\bibfnamefont {S.-S.}\ \bibnamefont {Lee}},
  \bibinfo {author} {\bibfnamefont {A.}~\bibnamefont {Weichselbaum}}, \bibinfo
  {author} {\bibfnamefont {G.}~\bibnamefont {Kotliar}},\ and\ \bibinfo {author}
  {\bibfnamefont {J.}~\bibnamefont {von Delft}},\ }\bibfield  {title} {\bibinfo
  {title} {{To be published}},\ }\href
  {https://meetings.aps.org/Meeting/MAR22/Session/K63.13} {\  (\bibinfo {year}
  {2022})}\BibitemShut {NoStop}%
\bibitem [{\citenamefont {Fye}(1994)}]{Fye}%
  \BibitemOpen
  \bibfield  {author} {\bibinfo {author} {\bibfnamefont {R.~M.}\ \bibnamefont
  {Fye}},\ }\bibfield  {title} {\bibinfo {title} {{Anomalous fixed point
  behavior'' of two Kondo impurities: A reexamination}},\ }\href
  {https://doi.org/10.1103/PhysRevLett.72.916} {\bibfield  {journal} {\bibinfo
  {journal} {Phys. Rev. Lett.}\ }\textbf {\bibinfo {volume} {72}},\ \bibinfo
  {pages} {916} (\bibinfo {year} {1994})}\BibitemShut {NoStop}%
\bibitem [{\citenamefont {Affleck}\ \emph {et~al.}(1995)\citenamefont
  {Affleck}, \citenamefont {Ludwig},\ and\ \citenamefont {Jones}}]{Affleck}%
  \BibitemOpen
  \bibfield  {author} {\bibinfo {author} {\bibfnamefont {I.}~\bibnamefont
  {Affleck}}, \bibinfo {author} {\bibfnamefont {A.~W.~W.}\ \bibnamefont
  {Ludwig}},\ and\ \bibinfo {author} {\bibfnamefont {B.~A.}\ \bibnamefont
  {Jones}},\ }\bibfield  {title} {\bibinfo {title} {{Conformal-field-theory
  approach to the two-impurity Kondo problem: Comparison with numerical
  renormalization-group results}},\ }\href
  {https://doi.org/10.1103/PhysRevB.52.9528} {\bibfield  {journal} {\bibinfo
  {journal} {Phys. Rev. B}\ }\textbf {\bibinfo {volume} {52}},\ \bibinfo
  {pages} {9528} (\bibinfo {year} {1995})}\BibitemShut {NoStop}%
\bibitem [{\citenamefont {Silva}\ \emph {et~al.}(1996)\citenamefont {Silva},
  \citenamefont {Lima}, \citenamefont {Oliveira}, \citenamefont {Mello},
  \citenamefont {Oliveira},\ and\ \citenamefont {Wilkins}}]{Silva1996Jan}%
  \BibitemOpen
  \bibfield  {author} {\bibinfo {author} {\bibfnamefont {J.~B.}\ \bibnamefont
  {Silva}}, \bibinfo {author} {\bibfnamefont {W.~L.~C.}\ \bibnamefont {Lima}},
  \bibinfo {author} {\bibfnamefont {W.~C.}\ \bibnamefont {Oliveira}}, \bibinfo
  {author} {\bibfnamefont {J.~L.~N.}\ \bibnamefont {Mello}}, \bibinfo {author}
  {\bibfnamefont {L.~N.}\ \bibnamefont {Oliveira}},\ and\ \bibinfo {author}
  {\bibfnamefont {J.~W.}\ \bibnamefont {Wilkins}},\ }\bibfield  {title}
  {\bibinfo {title} {{Particle-Hole Asymmetry in the Two-Impurity Kondo
  Model}},\ }\href {https://doi.org/10.1103/PhysRevLett.76.275} {\bibfield
  {journal} {\bibinfo  {journal} {Phys. Rev. Lett.}\ }\textbf {\bibinfo
  {volume} {76}},\ \bibinfo {pages} {275} (\bibinfo {year} {1996})}\BibitemShut
  {NoStop}%
\bibitem [{\citenamefont {Eickhoff}\ \emph {et~al.}(2018)\citenamefont
  {Eickhoff}, \citenamefont {Lechtenberg},\ and\ \citenamefont
  {Anders}}]{Fabian}%
  \BibitemOpen
  \bibfield  {author} {\bibinfo {author} {\bibfnamefont {F.}~\bibnamefont
  {Eickhoff}}, \bibinfo {author} {\bibfnamefont {B.}~\bibnamefont
  {Lechtenberg}},\ and\ \bibinfo {author} {\bibfnamefont {F.~B.}\ \bibnamefont
  {Anders}},\ }\bibfield  {title} {\bibinfo {title} {{Effective low-energy
  description of the two-impurity Anderson model: RKKY interaction and quantum
  criticality}},\ }\href {https://doi.org/10.1103/PhysRevB.98.115103}
  {\bibfield  {journal} {\bibinfo  {journal} {Phys. Rev. B}\ }\textbf {\bibinfo
  {volume} {98}},\ \bibinfo {pages} {115103} (\bibinfo {year}
  {2018})}\BibitemShut {NoStop}%
\bibitem [{\citenamefont {Zar{\ifmmode\acute{a}\else\'{a}\fi}nd}\ \emph
  {et~al.}(2006)\citenamefont {Zar{\ifmmode\acute{a}\else\'{a}\fi}nd},
  \citenamefont {Chung}, \citenamefont {Simon},\ and\ \citenamefont
  {Vojta}}]{Zarand2006Oct}%
  \BibitemOpen
  \bibfield  {author} {\bibinfo {author} {\bibfnamefont {G.}~\bibnamefont
  {Zar{\ifmmode\acute{a}\else\'{a}\fi}nd}}, \bibinfo {author} {\bibfnamefont
  {C.-H.}\ \bibnamefont {Chung}}, \bibinfo {author} {\bibfnamefont
  {P.}~\bibnamefont {Simon}},\ and\ \bibinfo {author} {\bibfnamefont
  {M.}~\bibnamefont {Vojta}},\ }\bibfield  {title} {\bibinfo {title} {{Quantum
  Criticality in a Double-Quantum-Dot System}},\ }\href
  {https://doi.org/10.1103/PhysRevLett.97.166802} {\bibfield  {journal}
  {\bibinfo  {journal} {Phys. Rev. Lett.}\ }\textbf {\bibinfo {volume} {97}},\
  \bibinfo {pages} {166802} (\bibinfo {year} {2006})}\BibitemShut {NoStop}%
\bibitem [{\citenamefont
  {Nozi{\ifmmode\grave{e}\else\`{e}\fi}res}(1985)}]{Nozieres1985}%
  \BibitemOpen
  \bibfield  {author} {\bibinfo {author} {\bibfnamefont {P.}~\bibnamefont
  {Nozi{\ifmmode\grave{e}\else\`{e}\fi}res}},\ }\bibfield  {title} {\bibinfo
  {title} {{Impuret{\ifmmode\acute{e}\else\'{e}\fi}s
  magn{\ifmmode\acute{e}\else\'{e}\fi}tiques et effet Kondo}},\ }\href
  {https://doi.org/10.1051/anphys:0198500100101900} {\bibfield  {journal}
  {\bibinfo  {journal} {Ann. Phys. Fr.}\ }\textbf {\bibinfo {volume} {10}},\
  \bibinfo {pages} {19} (\bibinfo {year} {1985})}\BibitemShut {NoStop}%
\bibitem [{\citenamefont
  {Nozi{\ifmmode\grave{e}\else\`{e}\fi}res}(1998)}]{Nozieres1998}%
  \BibitemOpen
  \bibfield  {author} {\bibinfo {author} {\bibfnamefont {{\relax
  Ph}.}~\bibnamefont {Nozi{\ifmmode\grave{e}\else\`{e}\fi}res}},\ }\bibfield
  {title} {\bibinfo {title} {{Some comments on Kondo lattices and the Mott
  transition}},\ }\href {https://doi.org/10.1007/s100510050571} {\bibfield
  {journal} {\bibinfo  {journal} {Eur. Phys. J. B}\ }\textbf {\bibinfo {volume}
  {6}},\ \bibinfo {pages} {447} (\bibinfo {year} {1998})}\BibitemShut {NoStop}%
\bibitem [{\citenamefont {Eickhoff}\ and\ \citenamefont
  {Anders}(2020)}]{Eickhoff2020Nov}%
  \BibitemOpen
  \bibfield  {author} {\bibinfo {author} {\bibfnamefont {F.}~\bibnamefont
  {Eickhoff}}\ and\ \bibinfo {author} {\bibfnamefont {F.~B.}\ \bibnamefont
  {Anders}},\ }\bibfield  {title} {\bibinfo {title} {{Strongly correlated
  multi-impurity models: The crossover from a single-impurity problem to
  lattice models}},\ }\href {https://doi.org/10.1103/PhysRevB.102.205132}
  {\bibfield  {journal} {\bibinfo  {journal} {Phys. Rev. B}\ }\textbf {\bibinfo
  {volume} {102}},\ \bibinfo {pages} {205132} (\bibinfo {year}
  {2020})}\BibitemShut {NoStop}%
\bibitem [{\citenamefont {Si}(2010)}]{Si2010}%
  \BibitemOpen
  \bibfield  {author} {\bibinfo {author} {\bibfnamefont {Q.}~\bibnamefont
  {Si}},\ }\bibfield  {title} {\bibinfo {title} {Quantum criticality and global
  phase diagram of magnetic heavy fermions},\ }\href
  {https://doi.org/https://doi.org/10.1002/pssb.200983082} {\bibfield
  {journal} {\bibinfo  {journal} {Phys. Stat. Solidi (b)}\ }\textbf {\bibinfo
  {volume} {247}},\ \bibinfo {pages} {476} (\bibinfo {year}
  {2010})}\BibitemShut {NoStop}%
\bibitem [{\citenamefont {Nakatsuji}\ \emph {et~al.}(2006)\citenamefont
  {Nakatsuji}, \citenamefont {Machida}, \citenamefont {Maeno}, \citenamefont
  {Tayama}, \citenamefont {Sakakibara}, \citenamefont {Duijn}, \citenamefont
  {Balicas}, \citenamefont {Millican}, \citenamefont {Macaluso},\ and\
  \citenamefont {Chan}}]{Nakatsuji2006Mar}%
  \BibitemOpen
  \bibfield  {author} {\bibinfo {author} {\bibfnamefont {S.}~\bibnamefont
  {Nakatsuji}}, \bibinfo {author} {\bibfnamefont {Y.}~\bibnamefont {Machida}},
  \bibinfo {author} {\bibfnamefont {Y.}~\bibnamefont {Maeno}}, \bibinfo
  {author} {\bibfnamefont {T.}~\bibnamefont {Tayama}}, \bibinfo {author}
  {\bibfnamefont {T.}~\bibnamefont {Sakakibara}}, \bibinfo {author}
  {\bibfnamefont {J.~v.}\ \bibnamefont {Duijn}}, \bibinfo {author}
  {\bibfnamefont {L.}~\bibnamefont {Balicas}}, \bibinfo {author} {\bibfnamefont
  {J.~N.}\ \bibnamefont {Millican}}, \bibinfo {author} {\bibfnamefont {R.~T.}\
  \bibnamefont {Macaluso}},\ and\ \bibinfo {author} {\bibfnamefont {J.~Y.}\
  \bibnamefont {Chan}},\ }\bibfield  {title} {\bibinfo {title} {{Metallic
  Spin-Liquid Behavior of the Geometrically Frustrated Kondo Lattice
  ${\mathrm{Pr}}_{2}{\mathrm{Ir}}_{2}{\mathrm{O}}_{7}$}},\ }\href
  {https://doi.org/10.1103/PhysRevLett.96.087204} {\bibfield  {journal}
  {\bibinfo  {journal} {Phys. Rev. Lett.}\ }\textbf {\bibinfo {volume} {96}},\
  \bibinfo {pages} {087204} (\bibinfo {year} {2006})}\BibitemShut {NoStop}%
\bibitem [{\citenamefont {Friedemann}\ \emph {et~al.}(2009)\citenamefont
  {Friedemann}, \citenamefont {Westerkamp}, \citenamefont {Brando},
  \citenamefont {Oeschler}, \citenamefont {Wirth}, \citenamefont {Gegenwart},
  \citenamefont {Krellner}, \citenamefont {Geibel},\ and\ \citenamefont
  {Steglich}}]{Friedemann2009Jul}%
  \BibitemOpen
  \bibfield  {author} {\bibinfo {author} {\bibfnamefont {S.}~\bibnamefont
  {Friedemann}}, \bibinfo {author} {\bibfnamefont {T.}~\bibnamefont
  {Westerkamp}}, \bibinfo {author} {\bibfnamefont {M.}~\bibnamefont {Brando}},
  \bibinfo {author} {\bibfnamefont {N.}~\bibnamefont {Oeschler}}, \bibinfo
  {author} {\bibfnamefont {S.}~\bibnamefont {Wirth}}, \bibinfo {author}
  {\bibfnamefont {P.}~\bibnamefont {Gegenwart}}, \bibinfo {author}
  {\bibfnamefont {C.}~\bibnamefont {Krellner}}, \bibinfo {author}
  {\bibfnamefont {C.}~\bibnamefont {Geibel}},\ and\ \bibinfo {author}
  {\bibfnamefont {F.}~\bibnamefont {Steglich}},\ }\bibfield  {title} {\bibinfo
  {title} {{Detaching the antiferromagnetic quantum critical point from the
  Fermi-surface reconstruction in YbRh2Si2}},\ }\href
  {https://doi.org/10.1038/nphys1299} {\bibfield  {journal} {\bibinfo
  {journal} {Nat. Phys.}\ }\textbf {\bibinfo {volume} {5}},\ \bibinfo {pages}
  {465} (\bibinfo {year} {2009})}\BibitemShut {NoStop}%
\bibitem [{\citenamefont {Lucas}\ \emph {et~al.}(2017)\citenamefont {Lucas},
  \citenamefont {Grube}, \citenamefont {Huang}, \citenamefont {Sakai},
  \citenamefont {Wunderlich}, \citenamefont {Green}, \citenamefont {Wosnitza},
  \citenamefont {Fritsch}, \citenamefont {Gegenwart}, \citenamefont
  {Stockert},\ and\ \citenamefont
  {V.~L{\ifmmode\ddot{o}\else\"{o}\fi}hneysen}}]{Lucas2017Mar}%
  \BibitemOpen
  \bibfield  {author} {\bibinfo {author} {\bibfnamefont {S.}~\bibnamefont
  {Lucas}}, \bibinfo {author} {\bibfnamefont {K.}~\bibnamefont {Grube}},
  \bibinfo {author} {\bibfnamefont {C.-L.}\ \bibnamefont {Huang}}, \bibinfo
  {author} {\bibfnamefont {A.}~\bibnamefont {Sakai}}, \bibinfo {author}
  {\bibfnamefont {S.}~\bibnamefont {Wunderlich}}, \bibinfo {author}
  {\bibfnamefont {E.~L.}\ \bibnamefont {Green}}, \bibinfo {author}
  {\bibfnamefont {J.}~\bibnamefont {Wosnitza}}, \bibinfo {author}
  {\bibfnamefont {V.}~\bibnamefont {Fritsch}}, \bibinfo {author} {\bibfnamefont
  {P.}~\bibnamefont {Gegenwart}}, \bibinfo {author} {\bibfnamefont
  {O.}~\bibnamefont {Stockert}},\ and\ \bibinfo {author} {\bibfnamefont
  {H.}~\bibnamefont {V.~L{\ifmmode\ddot{o}\else\"{o}\fi}hneysen}},\ }\bibfield
  {title} {\bibinfo {title} {{Entropy Evolution in the Magnetic Phases of
  Partially Frustrated CePdAl}},\ }\href
  {https://doi.org/10.1103/PhysRevLett.118.107204} {\bibfield  {journal}
  {\bibinfo  {journal} {Phys. Rev. Lett.}\ }\textbf {\bibinfo {volume} {118}},\
  \bibinfo {pages} {107204} (\bibinfo {year} {2017})}\BibitemShut {NoStop}%
\bibitem [{\citenamefont {Zhao}\ \emph {et~al.}(2019)\citenamefont {Zhao},
  \citenamefont {Zhang}, \citenamefont {Lyu}, \citenamefont {Bachus},
  \citenamefont {Tokiwa}, \citenamefont {Gegenwart}, \citenamefont {Zhang},
  \citenamefont {Cheng}, \citenamefont {Yang}, \citenamefont {Chen},
  \citenamefont {Isikawa}, \citenamefont {Si}, \citenamefont {Steglich},\ and\
  \citenamefont {Sun}}]{Zhao2019Dec}%
  \BibitemOpen
  \bibfield  {author} {\bibinfo {author} {\bibfnamefont {H.}~\bibnamefont
  {Zhao}}, \bibinfo {author} {\bibfnamefont {J.}~\bibnamefont {Zhang}},
  \bibinfo {author} {\bibfnamefont {M.}~\bibnamefont {Lyu}}, \bibinfo {author}
  {\bibfnamefont {S.}~\bibnamefont {Bachus}}, \bibinfo {author} {\bibfnamefont
  {Y.}~\bibnamefont {Tokiwa}}, \bibinfo {author} {\bibfnamefont
  {P.}~\bibnamefont {Gegenwart}}, \bibinfo {author} {\bibfnamefont
  {S.}~\bibnamefont {Zhang}}, \bibinfo {author} {\bibfnamefont
  {J.}~\bibnamefont {Cheng}}, \bibinfo {author} {\bibfnamefont {Y.-f.}\
  \bibnamefont {Yang}}, \bibinfo {author} {\bibfnamefont {G.}~\bibnamefont
  {Chen}}, \bibinfo {author} {\bibfnamefont {Y.}~\bibnamefont {Isikawa}},
  \bibinfo {author} {\bibfnamefont {Q.}~\bibnamefont {Si}}, \bibinfo {author}
  {\bibfnamefont {F.}~\bibnamefont {Steglich}},\ and\ \bibinfo {author}
  {\bibfnamefont {P.}~\bibnamefont {Sun}},\ }\bibfield  {title} {\bibinfo
  {title} {{Quantum-critical phase from frustrated magnetism in a strongly
  correlated metal}},\ }\href {https://doi.org/10.1038/s41567-019-0666-6}
  {\bibfield  {journal} {\bibinfo  {journal} {Nat. Phys.}\ }\textbf {\bibinfo
  {volume} {15}},\ \bibinfo {pages} {1261} (\bibinfo {year}
  {2019})}\BibitemShut {NoStop}%
\bibitem [{\citenamefont {Majumder}\ \emph {et~al.}(2022)\citenamefont
  {Majumder}, \citenamefont {Gupta}, \citenamefont {Luetkens}, \citenamefont
  {Khasanov}, \citenamefont {Stockert}, \citenamefont {Gegenwart},\ and\
  \citenamefont {Fritsch}}]{Majumder2022May}%
  \BibitemOpen
  \bibfield  {author} {\bibinfo {author} {\bibfnamefont {M.}~\bibnamefont
  {Majumder}}, \bibinfo {author} {\bibfnamefont {R.}~\bibnamefont {Gupta}},
  \bibinfo {author} {\bibfnamefont {H.}~\bibnamefont {Luetkens}}, \bibinfo
  {author} {\bibfnamefont {R.}~\bibnamefont {Khasanov}}, \bibinfo {author}
  {\bibfnamefont {O.}~\bibnamefont {Stockert}}, \bibinfo {author}
  {\bibfnamefont {P.}~\bibnamefont {Gegenwart}},\ and\ \bibinfo {author}
  {\bibfnamefont {V.}~\bibnamefont {Fritsch}},\ }\bibfield  {title} {\bibinfo
  {title} {{Spin-liquid signatures in the quantum critical regime of
  pressurized CePdAl}},\ }\href {https://doi.org/10.1103/PhysRevB.105.L180402}
  {\bibfield  {journal} {\bibinfo  {journal} {Phys. Rev. B}\ }\textbf {\bibinfo
  {volume} {105}},\ \bibinfo {pages} {L180402} (\bibinfo {year}
  {2022})}\BibitemShut {NoStop}%
\bibitem [{\citenamefont {Tripathi}\ \emph {et~al.}(2022)\citenamefont
  {Tripathi}, \citenamefont {Adroja}, \citenamefont {Ritter}, \citenamefont
  {Sharma}, \citenamefont {Yang}, \citenamefont {Hillier}, \citenamefont
  {Koza}, \citenamefont {Demmel}, \citenamefont {Sundaresan}, \citenamefont
  {Langridge}, \citenamefont {Higemoto}, \citenamefont {Ito}, \citenamefont
  {Strydom}, \citenamefont {Stenning}, \citenamefont {Bhattacharyya},
  \citenamefont {Keen}, \citenamefont {Walker}, \citenamefont {Perry},
  \citenamefont {Pratt}, \citenamefont {Si},\ and\ \citenamefont
  {Takabatake}}]{Tripathi2022Aug}%
  \BibitemOpen
  \bibfield  {author} {\bibinfo {author} {\bibfnamefont {R.}~\bibnamefont
  {Tripathi}}, \bibinfo {author} {\bibfnamefont {D.~T.}\ \bibnamefont
  {Adroja}}, \bibinfo {author} {\bibfnamefont {C.}~\bibnamefont {Ritter}},
  \bibinfo {author} {\bibfnamefont {S.}~\bibnamefont {Sharma}}, \bibinfo
  {author} {\bibfnamefont {C.}~\bibnamefont {Yang}}, \bibinfo {author}
  {\bibfnamefont {A.~D.}\ \bibnamefont {Hillier}}, \bibinfo {author}
  {\bibfnamefont {M.~M.}\ \bibnamefont {Koza}}, \bibinfo {author}
  {\bibfnamefont {F.}~\bibnamefont {Demmel}}, \bibinfo {author} {\bibfnamefont
  {A.}~\bibnamefont {Sundaresan}}, \bibinfo {author} {\bibfnamefont
  {S.}~\bibnamefont {Langridge}}, \bibinfo {author} {\bibfnamefont
  {W.}~\bibnamefont {Higemoto}}, \bibinfo {author} {\bibfnamefont {T.~U.}\
  \bibnamefont {Ito}}, \bibinfo {author} {\bibfnamefont {A.~M.}\ \bibnamefont
  {Strydom}}, \bibinfo {author} {\bibfnamefont {G.~B.~G.}\ \bibnamefont
  {Stenning}}, \bibinfo {author} {\bibfnamefont {A.}~\bibnamefont
  {Bhattacharyya}}, \bibinfo {author} {\bibfnamefont {D.}~\bibnamefont {Keen}},
  \bibinfo {author} {\bibfnamefont {H.~C.}\ \bibnamefont {Walker}}, \bibinfo
  {author} {\bibfnamefont {R.~S.}\ \bibnamefont {Perry}}, \bibinfo {author}
  {\bibfnamefont {F.}~\bibnamefont {Pratt}}, \bibinfo {author} {\bibfnamefont
  {Q.}~\bibnamefont {Si}},\ and\ \bibinfo {author} {\bibfnamefont
  {T.}~\bibnamefont {Takabatake}},\ }\bibfield  {title} {\bibinfo {title}
  {{Quantum critical spin-liquid-like behavior in S = 1/2 quasikagome lattice
  CeRh1-xPdxSn investigated using muon spin relaxation and neutron
  scattering}},\ }\bibfield  {journal} {\bibinfo  {journal} {arXiv}\ }\href
  {https://doi.org/10.48550/arXiv.2208.03148} {10.48550/arXiv.2208.03148}
  (\bibinfo {year} {2022}),\ \Eprint {https://arxiv.org/abs/2208.03148}
  {2208.03148} \BibitemShut {NoStop}%
\bibitem [{\citenamefont {Nobukane}\ \emph {et~al.}(2020)\citenamefont
  {Nobukane}, \citenamefont {Tabata}, \citenamefont {Kurosawa}, \citenamefont
  {Sakabe},\ and\ \citenamefont {Tanda}}]{Nobukane2020Jan}%
  \BibitemOpen
  \bibfield  {author} {\bibinfo {author} {\bibfnamefont {H.}~\bibnamefont
  {Nobukane}}, \bibinfo {author} {\bibfnamefont {Y.}~\bibnamefont {Tabata}},
  \bibinfo {author} {\bibfnamefont {T.}~\bibnamefont {Kurosawa}}, \bibinfo
  {author} {\bibfnamefont {D.}~\bibnamefont {Sakabe}},\ and\ \bibinfo {author}
  {\bibfnamefont {S.}~\bibnamefont {Tanda}},\ }\bibfield  {title} {\bibinfo
  {title} {{Coexistence of the Kondo effect and spin glass physics in Fe-doped
  NbS2}},\ }\href {https://doi.org/10.1088/1361-648x/ab622a} {\bibfield
  {journal} {\bibinfo  {journal} {J. Phys.: Condens. Matter}\ }\textbf
  {\bibinfo {volume} {32}},\ \bibinfo {pages} {165803} (\bibinfo {year}
  {2020})}\BibitemShut {NoStop}%
\bibitem [{\citenamefont {Lacroix}\ \emph {et~al.}(1996)\citenamefont
  {Lacroix}, \citenamefont {Canals},\ and\ \citenamefont
  {N{\ifmmode\acute{u}\else\'{u}\fi}{\ifmmode\tilde{n}\else\~{n}\fi}ez-Regueiro}}]{Lacroix1996Dec}%
  \BibitemOpen
  \bibfield  {author} {\bibinfo {author} {\bibfnamefont {C.}~\bibnamefont
  {Lacroix}}, \bibinfo {author} {\bibfnamefont {B.}~\bibnamefont {Canals}},\
  and\ \bibinfo {author} {\bibfnamefont {M.~D.}\ \bibnamefont
  {N{\ifmmode\acute{u}\else\'{u}\fi}{\ifmmode\tilde{n}\else\~{n}\fi}ez-Regueiro}},\
  }\bibfield  {title} {\bibinfo {title} {{Kondo Screening and Magnetic Ordering
  in Frustrated ${\mathrm{UNi}}_{4}\mathrm{B}$}},\ }\href
  {https://doi.org/10.1103/PhysRevLett.77.5126} {\bibfield  {journal} {\bibinfo
   {journal} {Phys. Rev. Lett.}\ }\textbf {\bibinfo {volume} {77}},\ \bibinfo
  {pages} {5126} (\bibinfo {year} {1996})}\BibitemShut {NoStop}%
\bibitem [{\citenamefont {Li}\ \emph {et~al.}(2010)\citenamefont {Li},
  \citenamefont {Zhang},\ and\ \citenamefont {Yu}}]{Li2010Mar}%
  \BibitemOpen
  \bibfield  {author} {\bibinfo {author} {\bibfnamefont {G.-B.}\ \bibnamefont
  {Li}}, \bibinfo {author} {\bibfnamefont {G.-M.}\ \bibnamefont {Zhang}},\ and\
  \bibinfo {author} {\bibfnamefont {L.}~\bibnamefont {Yu}},\ }\bibfield
  {title} {\bibinfo {title} {{Kondo screening coexisting with ferromagnetic
  order as a possible ground state for Kondo lattice systems}},\ }\href
  {https://doi.org/10.1103/PhysRevB.81.094420} {\bibfield  {journal} {\bibinfo
  {journal} {Phys. Rev. B}\ }\textbf {\bibinfo {volume} {81}},\ \bibinfo
  {pages} {094420} (\bibinfo {year} {2010})}\BibitemShut {NoStop}%
\bibitem [{\citenamefont {Motome}\ \emph {et~al.}(2010)\citenamefont {Motome},
  \citenamefont {Nakamikawa}, \citenamefont {Yamaji},\ and\ \citenamefont
  {Udagawa}}]{Motome2010Jul}%
  \BibitemOpen
  \bibfield  {author} {\bibinfo {author} {\bibfnamefont {Y.}~\bibnamefont
  {Motome}}, \bibinfo {author} {\bibfnamefont {K.}~\bibnamefont {Nakamikawa}},
  \bibinfo {author} {\bibfnamefont {Y.}~\bibnamefont {Yamaji}},\ and\ \bibinfo
  {author} {\bibfnamefont {M.}~\bibnamefont {Udagawa}},\ }\bibfield  {title}
  {\bibinfo {title} {{Partial Kondo Screening in Frustrated Kondo Lattice
  Systems}},\ }\href {https://doi.org/10.1103/PhysRevLett.105.036403}
  {\bibfield  {journal} {\bibinfo  {journal} {Phys. Rev. Lett.}\ }\textbf
  {\bibinfo {volume} {105}},\ \bibinfo {pages} {036403} (\bibinfo {year}
  {2010})}\BibitemShut {NoStop}%
\bibitem [{\citenamefont {Bernhard}\ and\ \citenamefont
  {Lacroix}(2015)}]{Bernhard2015Sep}%
  \BibitemOpen
  \bibfield  {author} {\bibinfo {author} {\bibfnamefont {B.~H.}\ \bibnamefont
  {Bernhard}}\ and\ \bibinfo {author} {\bibfnamefont {C.}~\bibnamefont
  {Lacroix}},\ }\bibfield  {title} {\bibinfo {title} {{Coexistence of magnetic
  order and Kondo effect in the Kondo-Heisenberg model}},\ }\href
  {https://doi.org/10.1103/PhysRevB.92.094401} {\bibfield  {journal} {\bibinfo
  {journal} {Phys. Rev. B}\ }\textbf {\bibinfo {volume} {92}},\ \bibinfo
  {pages} {094401} (\bibinfo {year} {2015})}\BibitemShut {NoStop}%
\bibitem [{\citenamefont {Sato}\ \emph {et~al.}(2018)\citenamefont {Sato},
  \citenamefont {Assaad},\ and\ \citenamefont {Grover}}]{Sato2018Mar}%
  \BibitemOpen
  \bibfield  {author} {\bibinfo {author} {\bibfnamefont {T.}~\bibnamefont
  {Sato}}, \bibinfo {author} {\bibfnamefont {F.~F.}\ \bibnamefont {Assaad}},\
  and\ \bibinfo {author} {\bibfnamefont {T.}~\bibnamefont {Grover}},\
  }\bibfield  {title} {\bibinfo {title} {{Quantum Monte Carlo Simulation of
  Frustrated Kondo Lattice Models}},\ }\href
  {https://doi.org/10.1103/PhysRevLett.120.107201} {\bibfield  {journal}
  {\bibinfo  {journal} {Phys. Rev. Lett.}\ }\textbf {\bibinfo {volume} {120}},\
  \bibinfo {pages} {107201} (\bibinfo {year} {2018})}\BibitemShut {NoStop}%
\bibitem [{\citenamefont {W{\ifmmode\acute{o}\else\'{o}\fi}jcik}\ \emph
  {et~al.}(2020)\citenamefont {W{\ifmmode\acute{o}\else\'{o}\fi}jcik},
  \citenamefont {Weymann},\ and\ \citenamefont {Kroha}}]{KWIWJK_3QD}%
  \BibitemOpen
  \bibfield  {author} {\bibinfo {author} {\bibfnamefont {K.~P.}\ \bibnamefont
  {W{\ifmmode\acute{o}\else\'{o}\fi}jcik}}, \bibinfo {author} {\bibfnamefont
  {I.}~\bibnamefont {Weymann}},\ and\ \bibinfo {author} {\bibfnamefont
  {J.}~\bibnamefont {Kroha}},\ }\bibfield  {title} {\bibinfo {title} {{Magnetic
  Kondo regimes in a frustrated half-filled trimer}},\ }\href
  {https://doi.org/10.1103/PhysRevB.102.045144} {\bibfield  {journal} {\bibinfo
   {journal} {Phys. Rev. B}\ }\textbf {\bibinfo {volume} {102}},\ \bibinfo
  {pages} {045144} (\bibinfo {year} {2020})}\BibitemShut {NoStop}%
\bibitem [{\citenamefont {Ke{\ss}ler}\ and\ \citenamefont
  {Eder}(2020)}]{Kessler2020Dec}%
  \BibitemOpen
  \bibfield  {author} {\bibinfo {author} {\bibfnamefont {M.}~\bibnamefont
  {Ke{\ss}ler}}\ and\ \bibinfo {author} {\bibfnamefont {R.}~\bibnamefont
  {Eder}},\ }\bibfield  {title} {\bibinfo {title} {{Magnetic phases of the
  triangular Kondo lattice}},\ }\href
  {https://doi.org/10.1103/PhysRevB.102.235125} {\bibfield  {journal} {\bibinfo
   {journal} {Phys. Rev. B}\ }\textbf {\bibinfo {volume} {102}},\ \bibinfo
  {pages} {235125} (\bibinfo {year} {2020})}\BibitemShut {NoStop}%
\bibitem [{\citenamefont {W\'{o}jcik}\ and\ \citenamefont
  {Kroha}()}]{KWJK_2imp}%
  \BibitemOpen
  \bibfield  {author} {\bibinfo {author} {\bibfnamefont {K.~P.}\ \bibnamefont
  {W\'{o}jcik}}\ and\ \bibinfo {author} {\bibfnamefont {J.}~\bibnamefont
  {Kroha}},\ }\bibfield  {title} {\bibinfo {title} {{Quantum spin liquid in an
  RKKY-coupled two-impurity Kondo system}},\ }\href
  {https://arxiv.org/abs/2106.07519} {\bibinfo  {journal} {arXiv 2106.07519v2
  (2022)}\ }\BibitemShut {NoStop}%
\bibitem [{\citenamefont {Coleman}\ and\ \citenamefont
  {Andrei}(1989)}]{Coleman1989Jul}%
  \BibitemOpen
\bibfield  {journal} {  }\bibfield  {author} {\bibinfo {author} {\bibfnamefont
  {P.}~\bibnamefont {Coleman}}\ and\ \bibinfo {author} {\bibfnamefont
  {N.}~\bibnamefont {Andrei}},\ }\bibfield  {title} {\bibinfo {title}
  {{Kondo-stabilised spin liquids and heavy fermion superconductivity}},\
  }\href {https://doi.org/10.1088/0953-8984/1/26/003} {\bibfield  {journal}
  {\bibinfo  {journal} {J. Phys.: Condens. Matter}\ }\textbf {\bibinfo {volume}
  {1}},\ \bibinfo {pages} {4057} (\bibinfo {year} {1989})}\BibitemShut
  {NoStop}%
\bibitem [{\citenamefont {Andrei}\ and\ \citenamefont
  {Coleman}(1989)}]{Andrei1989Jan}%
  \BibitemOpen
  \bibfield  {author} {\bibinfo {author} {\bibfnamefont {N.}~\bibnamefont
  {Andrei}}\ and\ \bibinfo {author} {\bibfnamefont {P.}~\bibnamefont
  {Coleman}},\ }\bibfield  {title} {\bibinfo {title} {{Cooper Instability in
  the Presence of a Spin Liquid}},\ }\href
  {https://doi.org/10.1103/PhysRevLett.62.595} {\bibfield  {journal} {\bibinfo
  {journal} {Phys. Rev. Lett.}\ }\textbf {\bibinfo {volume} {62}},\ \bibinfo
  {pages} {595} (\bibinfo {year} {1989})}\BibitemShut {NoStop}%
\bibitem [{\citenamefont {Peschke}\ \emph {et~al.}(2022)\citenamefont
  {Peschke}, \citenamefont {Ponsioen},\ and\ \citenamefont
  {Corboz}}]{Peschke2022Sep}%
  \BibitemOpen
  \bibfield  {author} {\bibinfo {author} {\bibfnamefont {M.}~\bibnamefont
  {Peschke}}, \bibinfo {author} {\bibfnamefont {B.}~\bibnamefont {Ponsioen}},\
  and\ \bibinfo {author} {\bibfnamefont {P.}~\bibnamefont {Corboz}},\
  }\bibfield  {title} {\bibinfo {title} {{Competing States in the
  Two-Dimensional Frustrated Kondo-Necklace Model}},\ }\href
  {10.48550/arXiv.2209.04231} {\bibfield  {journal} {\bibinfo  {journal}
  {arXiv:2209.04231}\ } (\bibinfo {year} {2022})}\BibitemShut {NoStop}%
\bibitem [{\citenamefont {Wilson}(1975)}]{WilsonNRG}%
  \BibitemOpen
  \bibfield  {author} {\bibinfo {author} {\bibfnamefont {K.~G.}\ \bibnamefont
  {Wilson}},\ }\bibfield  {title} {\bibinfo {title} {{The renormalization
  group: Critical phenomena and the Kondo problem}},\ }\href
  {https://doi.org/10.1103/RevModPhys.47.773} {\bibfield  {journal} {\bibinfo
  {journal} {Rev. Mod. Phys.}\ }\textbf {\bibinfo {volume} {47}},\ \bibinfo
  {pages} {773} (\bibinfo {year} {1975})}\BibitemShut {NoStop}%
\bibitem [{\citenamefont {Sela}\ and\ \citenamefont
  {Affleck}(2009)}]{SelaAffleck}%
  \BibitemOpen
  \bibfield  {author} {\bibinfo {author} {\bibfnamefont {E.}~\bibnamefont
  {Sela}}\ and\ \bibinfo {author} {\bibfnamefont {I.}~\bibnamefont {Affleck}},\
  }\bibfield  {title} {\bibinfo {title} {{Resonant Pair Tunneling in Double
  Quantum Dots}},\ }\href {https://doi.org/10.1103/PhysRevLett.103.087204}
  {\bibfield  {journal} {\bibinfo  {journal} {Phys. Rev. Lett.}\ }\textbf
  {\bibinfo {volume} {103}},\ \bibinfo {pages} {087204} (\bibinfo {year}
  {2009})}\BibitemShut {NoStop}%
\bibitem [{\citenamefont {Held}\ and\ \citenamefont
  {Vollhardt}(2000)}]{Held2000May}%
  \BibitemOpen
  \bibfield  {author} {\bibinfo {author} {\bibfnamefont {K.}~\bibnamefont
  {Held}}\ and\ \bibinfo {author} {\bibfnamefont {D.}~\bibnamefont
  {Vollhardt}},\ }\bibfield  {title} {\bibinfo {title} {{Electronic
  Correlations in Manganites}},\ }\href
  {https://doi.org/10.1103/PhysRevLett.84.5168} {\bibfield  {journal} {\bibinfo
   {journal} {Phys. Rev. Lett.}\ }\textbf {\bibinfo {volume} {84}},\ \bibinfo
  {pages} {5168} (\bibinfo {year} {2000})}\BibitemShut {NoStop}%
\bibitem [{\citenamefont {Hafez-Torbati}\ \emph {et~al.}(2021)\citenamefont
  {Hafez-Torbati}, \citenamefont {Bossini}, \citenamefont {Anders},\ and\
  \citenamefont {Uhrig}}]{Hafez-Torbati2021Dec}%
  \BibitemOpen
  \bibfield  {author} {\bibinfo {author} {\bibfnamefont {M.}~\bibnamefont
  {Hafez-Torbati}}, \bibinfo {author} {\bibfnamefont {D.}~\bibnamefont
  {Bossini}}, \bibinfo {author} {\bibfnamefont {F.~B.}\ \bibnamefont
  {Anders}},\ and\ \bibinfo {author} {\bibfnamefont {G.~S.}\ \bibnamefont
  {Uhrig}},\ }\bibfield  {title} {\bibinfo {title} {{Magnetic blue shift of
  Mott gaps enhanced by double exchange}},\ }\href
  {https://doi.org/10.1103/PhysRevResearch.3.043232} {\bibfield  {journal}
  {\bibinfo  {journal} {Phys. Rev. Res.}\ }\textbf {\bibinfo {volume} {3}},\
  \bibinfo {pages} {043232} (\bibinfo {year} {2021})}\BibitemShut {NoStop}%
\bibitem [{\citenamefont {Bulla}\ \emph {et~al.}(2008)\citenamefont {Bulla},
  \citenamefont {Costi},\ and\ \citenamefont {Pruschke}}]{NRG_RMP}%
  \BibitemOpen
  \bibfield  {author} {\bibinfo {author} {\bibfnamefont {R.}~\bibnamefont
  {Bulla}}, \bibinfo {author} {\bibfnamefont {T.~A.}\ \bibnamefont {Costi}},\
  and\ \bibinfo {author} {\bibfnamefont {T.}~\bibnamefont {Pruschke}},\
  }\bibfield  {title} {\bibinfo {title} {{Numerical renormalization group
  method for quantum impurity systems}},\ }\href
  {https://doi.org/10.1103/RevModPhys.80.395} {\bibfield  {journal} {\bibinfo
  {journal} {Rev. Mod. Phys.}\ }\textbf {\bibinfo {volume} {80}},\ \bibinfo
  {pages} {395} (\bibinfo {year} {2008})}\BibitemShut {NoStop}%
\bibitem [{\citenamefont {Legeza}\ \emph {et~al.}()\citenamefont {Legeza},
  \citenamefont {Moca}, \citenamefont {Toth}, \citenamefont {Weymann},\ and\
  \citenamefont {Zarand}}]{fnrg}%
  \BibitemOpen
  \bibfield  {author} {\bibinfo {author} {\bibfnamefont {O.}~\bibnamefont
  {Legeza}}, \bibinfo {author} {\bibfnamefont {C.~P.}\ \bibnamefont {Moca}},
  \bibinfo {author} {\bibfnamefont {A.~I.}\ \bibnamefont {Toth}}, \bibinfo
  {author} {\bibfnamefont {I.}~\bibnamefont {Weymann}},\ and\ \bibinfo {author}
  {\bibfnamefont {G.}~\bibnamefont {Zarand}},\ }\bibfield  {title} {\bibinfo
  {title} {{Manual for the Flexible DM-NRG code}},\ }\href
  {https://arxiv.org/abs/0809.3143} {\bibfield  {journal} {\bibinfo  {journal}
  {arXiv:0809.3143 (2008).}\ }}\bibinfo {note} {The code is available at
  \url{http://www.phy.bme.hu/~dmnrg/}}\BibitemShut {NoStop}%
\bibitem [{\citenamefont {Anders}\ and\ \citenamefont
  {Schiller}(2005)}]{AndersSchiller1}%
  \BibitemOpen
  \bibfield  {author} {\bibinfo {author} {\bibfnamefont {F.~B.}\ \bibnamefont
  {Anders}}\ and\ \bibinfo {author} {\bibfnamefont {A.}~\bibnamefont
  {Schiller}},\ }\bibfield  {title} {\bibinfo {title} {{Real-Time Dynamics in
  Quantum-Impurity Systems: A Time-Dependent Numerical Renormalization-Group
  Approach}},\ }\href {https://doi.org/10.1103/PhysRevLett.95.196801}
  {\bibfield  {journal} {\bibinfo  {journal} {Phys. Rev. Lett.}\ }\textbf
  {\bibinfo {volume} {95}},\ \bibinfo {pages} {196801} (\bibinfo {year}
  {2005})}\BibitemShut {NoStop}%
\bibitem [{\citenamefont {Weichselbaum}\ and\ \citenamefont {von
  Delft}(2007)}]{Weichselbaum}%
  \BibitemOpen
  \bibfield  {author} {\bibinfo {author} {\bibfnamefont {A.}~\bibnamefont
  {Weichselbaum}}\ and\ \bibinfo {author} {\bibfnamefont {J.}~\bibnamefont {von
  Delft}},\ }\bibfield  {title} {\bibinfo {title} {{Sum-Rule Conserving
  Spectral Functions from the Numerical Renormalization Group}},\ }\href
  {https://doi.org/10.1103/PhysRevLett.99.076402} {\bibfield  {journal}
  {\bibinfo  {journal} {Phys. Rev. Lett.}\ }\textbf {\bibinfo {volume} {99}},\
  \bibinfo {pages} {076402} (\bibinfo {year} {2007})}\BibitemShut {NoStop}%
\bibitem [{\citenamefont {Nejati}\ \emph {et~al.}(2017)\citenamefont {Nejati},
  \citenamefont {Ballmann},\ and\ \citenamefont {Kroha}}]{Hans}%
  \BibitemOpen
  \bibfield  {author} {\bibinfo {author} {\bibfnamefont {A.}~\bibnamefont
  {Nejati}}, \bibinfo {author} {\bibfnamefont {K.}~\bibnamefont {Ballmann}},\
  and\ \bibinfo {author} {\bibfnamefont {J.}~\bibnamefont {Kroha}},\ }\bibfield
   {title} {\bibinfo {title} {{Kondo Destruction in RKKY-Coupled Kondo Lattice
  and Multi-Impurity Systems}},\ }\href
  {https://doi.org/10.1103/PhysRevLett.118.117204} {\bibfield  {journal}
  {\bibinfo  {journal} {Phys. Rev. Lett.}\ }\textbf {\bibinfo {volume} {118}},\
  \bibinfo {pages} {117204} (\bibinfo {year} {2017})}\BibitemShut {NoStop}%
\bibitem [{\citenamefont {Si}\ \emph {et~al.}(2001)\citenamefont {Si},
  \citenamefont {Rabello}, \citenamefont {Ingersent},\ and\ \citenamefont
  {Smith}}]{Si2001Oct}%
  \BibitemOpen
  \bibfield  {author} {\bibinfo {author} {\bibfnamefont {Q.}~\bibnamefont
  {Si}}, \bibinfo {author} {\bibfnamefont {S.}~\bibnamefont {Rabello}},
  \bibinfo {author} {\bibfnamefont {K.}~\bibnamefont {Ingersent}},\ and\
  \bibinfo {author} {\bibfnamefont {J.~L.}\ \bibnamefont {Smith}},\ }\bibfield
  {title} {\bibinfo {title} {{Locally critical quantum phase transitions in
  strongly correlated metals}},\ }\href {https://doi.org/10.1038/35101507}
  {\bibfield  {journal} {\bibinfo  {journal} {Nature}\ }\textbf {\bibinfo
  {volume} {413}},\ \bibinfo {pages} {804} (\bibinfo {year}
  {2001})}\BibitemShut {NoStop}%
\bibitem [{\citenamefont {Coleman}\ \emph {et~al.}(2001)\citenamefont
  {Coleman}, \citenamefont {P{\ifmmode\acute{e}\else\'{e}\fi}pin},
  \citenamefont {Si},\ and\ \citenamefont {Ramazashvili}}]{Coleman2001Aug}%
  \BibitemOpen
  \bibfield  {author} {\bibinfo {author} {\bibfnamefont {P.}~\bibnamefont
  {Coleman}}, \bibinfo {author} {\bibfnamefont {C.}~\bibnamefont
  {P{\ifmmode\acute{e}\else\'{e}\fi}pin}}, \bibinfo {author} {\bibfnamefont
  {Q.}~\bibnamefont {Si}},\ and\ \bibinfo {author} {\bibfnamefont
  {R.}~\bibnamefont {Ramazashvili}},\ }\bibfield  {title} {\bibinfo {title}
  {{How do Fermi liquids get heavy and die?}},\ }\href
  {https://doi.org/10.1088/0953-8984/13/35/202} {\bibfield  {journal} {\bibinfo
   {journal} {J. Phys.: Condens. Matter}\ }\textbf {\bibinfo {volume} {13}},\
  \bibinfo {pages} {R723} (\bibinfo {year} {2001})}\BibitemShut {NoStop}%
\bibitem [{\citenamefont {Senthil}\ \emph {et~al.}(2004)\citenamefont
  {Senthil}, \citenamefont {Vojta},\ and\ \citenamefont
  {Sachdev}}]{Senthil2004Jan}%
  \BibitemOpen
  \bibfield  {author} {\bibinfo {author} {\bibfnamefont {T.}~\bibnamefont
  {Senthil}}, \bibinfo {author} {\bibfnamefont {M.}~\bibnamefont {Vojta}},\
  and\ \bibinfo {author} {\bibfnamefont {S.}~\bibnamefont {Sachdev}},\
  }\bibfield  {title} {\bibinfo {title} {{Weak magnetism and non-Fermi liquids
  near heavy-fermion critical points}},\ }\href
  {https://doi.org/10.1103/PhysRevB.69.035111} {\bibfield  {journal} {\bibinfo
  {journal} {Phys. Rev. B}\ }\textbf {\bibinfo {volume} {69}},\ \bibinfo
  {pages} {035111} (\bibinfo {year} {2004})}\BibitemShut {NoStop}%
\bibitem [{\citenamefont {Bulla}\ \emph {et~al.}(1997)\citenamefont {Bulla},
  \citenamefont {Pruschke},\ and\ \citenamefont {Hewson}}]{dosNRG}%
  \BibitemOpen
  \bibfield  {author} {\bibinfo {author} {\bibfnamefont {R.}~\bibnamefont
  {Bulla}}, \bibinfo {author} {\bibfnamefont {T.}~\bibnamefont {Pruschke}},\
  and\ \bibinfo {author} {\bibfnamefont {A.~C.}\ \bibnamefont {Hewson}},\
  }\bibfield  {title} {\bibinfo {title} {{Anderson impurity in pseudo-gap Fermi
  systems}},\ }\href {https://doi.org/10.1088/0953-8984/9/47/014} {\bibfield
  {journal} {\bibinfo  {journal} {J. Phys.: Condens. Matter}\ }\textbf
  {\bibinfo {volume} {9}},\ \bibinfo {pages} {10463} (\bibinfo {year}
  {1997})}\BibitemShut {NoStop}%
\bibitem [{\citenamefont {Fritz}\ and\ \citenamefont
  {Vojta}(2004)}]{Fritz2004Dec}%
  \BibitemOpen
  \bibfield  {author} {\bibinfo {author} {\bibfnamefont {L.}~\bibnamefont
  {Fritz}}\ and\ \bibinfo {author} {\bibfnamefont {M.}~\bibnamefont {Vojta}},\
  }\bibfield  {title} {\bibinfo {title} {{Phase transitions in the pseudogap
  Anderson and Kondo models: Critical dimensions, renormalization group, and
  local-moment criticality}},\ }\href
  {https://doi.org/10.1103/PhysRevB.70.214427} {\bibfield  {journal} {\bibinfo
  {journal} {Phys. Rev. B}\ }\textbf {\bibinfo {volume} {70}},\ \bibinfo
  {pages} {214427} (\bibinfo {year} {2004})}\BibitemShut {NoStop}%
\bibitem [{\citenamefont {Esat}\ \emph {et~al.}(2016)\citenamefont {Esat},
  \citenamefont {Lechtenberg}, \citenamefont {Deilmann}, \citenamefont
  {Wagner}, \citenamefont {Kr{\ifmmode\ddot{u}\else\"{u}\fi}ger}, \citenamefont
  {Temirov}, \citenamefont {Rohlfing}, \citenamefont {Anders},\ and\
  \citenamefont {Tautz}}]{AndersNatPhys}%
  \BibitemOpen
  \bibfield  {author} {\bibinfo {author} {\bibfnamefont {T.}~\bibnamefont
  {Esat}}, \bibinfo {author} {\bibfnamefont {B.}~\bibnamefont {Lechtenberg}},
  \bibinfo {author} {\bibfnamefont {T.}~\bibnamefont {Deilmann}}, \bibinfo
  {author} {\bibfnamefont {C.}~\bibnamefont {Wagner}}, \bibinfo {author}
  {\bibfnamefont {P.}~\bibnamefont {Kr{\ifmmode\ddot{u}\else\"{u}\fi}ger}},
  \bibinfo {author} {\bibfnamefont {R.}~\bibnamefont {Temirov}}, \bibinfo
  {author} {\bibfnamefont {M.}~\bibnamefont {Rohlfing}}, \bibinfo {author}
  {\bibfnamefont {F.~B.}\ \bibnamefont {Anders}},\ and\ \bibinfo {author}
  {\bibfnamefont {F.~S.}\ \bibnamefont {Tautz}},\ }\bibfield  {title} {\bibinfo
  {title} {{A chemically driven quantum phase transition in a two-molecule
  Kondo system}},\ }\href {https://doi.org/10.1038/nphys3737} {\bibfield
  {journal} {\bibinfo  {journal} {Nat. Phys.}\ }\textbf {\bibinfo {volume}
  {12}},\ \bibinfo {pages} {867} (\bibinfo {year} {2016})}\BibitemShut
  {NoStop}%
\bibitem [{\citenamefont {Georges}\ \emph {et~al.}(1996)\citenamefont
  {Georges}, \citenamefont {Kotliar}, \citenamefont {Krauth},\ and\
  \citenamefont {Rozenberg}}]{DMFT_RMP}%
  \BibitemOpen
  \bibfield  {author} {\bibinfo {author} {\bibfnamefont {A.}~\bibnamefont
  {Georges}}, \bibinfo {author} {\bibfnamefont {G.}~\bibnamefont {Kotliar}},
  \bibinfo {author} {\bibfnamefont {W.}~\bibnamefont {Krauth}},\ and\ \bibinfo
  {author} {\bibfnamefont {M.~J.}\ \bibnamefont {Rozenberg}},\ }\bibfield
  {title} {\bibinfo {title} {{Dynamical mean-field theory of strongly
  correlated fermion systems and the limit of infinite dimensions}},\ }\href
  {https://doi.org/10.1103/RevModPhys.68.13} {\bibfield  {journal} {\bibinfo
  {journal} {Rev. Mod. Phys.}\ }\textbf {\bibinfo {volume} {68}},\ \bibinfo
  {pages} {13} (\bibinfo {year} {1996})}\BibitemShut {NoStop}%
\bibitem [{\citenamefont {Millis}\ \emph {et~al.}(1990)\citenamefont {Millis},
  \citenamefont {Kotliar},\ and\ \citenamefont {Jones}}]{phi_book}%
  \BibitemOpen
  \bibfield  {author} {\bibinfo {author} {\bibfnamefont {A.~J.}\ \bibnamefont
  {Millis}}, \bibinfo {author} {\bibfnamefont {B.~G.}\ \bibnamefont
  {Kotliar}},\ and\ \bibinfo {author} {\bibfnamefont {B.~A.}\ \bibnamefont
  {Jones}},\ }\href@noop {} {\emph {\bibinfo {title} {{In: Z. Tesanovic (ed.),
  Many-Body Methods for Real Materials, pp.~159--166}}}}\ (\bibinfo
  {publisher} {Addison Wesley},\ \bibinfo {address} {Redwood City, CA},\
  \bibinfo {year} {1990})\BibitemShut {NoStop}%
\bibitem [{\citenamefont {Haldane}(1978)}]{Haldane}%
  \BibitemOpen
  \bibfield  {author} {\bibinfo {author} {\bibfnamefont {F.~D.~M.}\
  \bibnamefont {Haldane}},\ }\bibfield  {title} {\bibinfo {title} {{Scaling
  Theory of the Asymmetric Anderson Model}},\ }\href
  {https://doi.org/10.1103/PhysRevLett.40.416} {\bibfield  {journal} {\bibinfo
  {journal} {Phys. Rev. Lett.}\ }\textbf {\bibinfo {volume} {40}},\ \bibinfo
  {pages} {416} (\bibinfo {year} {1978})}\BibitemShut {NoStop}%
\bibitem [{\citenamefont {Hewson}(1997)}]{Hewson_book}%
  \BibitemOpen
  \bibfield  {author} {\bibinfo {author} {\bibfnamefont {A.~C.}\ \bibnamefont
  {Hewson}},\ }\href {https://doi.org/10.1017/CBO9780511470752} {\emph
  {\bibinfo {title} {{The Kondo problem to heavy fermions}}}}\ (\bibinfo
  {publisher} {Cambridge University Press},\ \bibinfo {address} {Cambridge},\
  \bibinfo {year} {1997})\BibitemShut {NoStop}%
\bibitem [{\citenamefont {Krishna-murthy}\ \emph {et~al.}(1980)\citenamefont
  {Krishna-murthy}, \citenamefont {Wilkins},\ and\ \citenamefont
  {Wilson}}]{KMWWb}%
  \BibitemOpen
  \bibfield  {author} {\bibinfo {author} {\bibfnamefont {H.~R.}\ \bibnamefont
  {Krishna-murthy}}, \bibinfo {author} {\bibfnamefont {J.~W.}\ \bibnamefont
  {Wilkins}},\ and\ \bibinfo {author} {\bibfnamefont {K.~G.}\ \bibnamefont
  {Wilson}},\ }\bibfield  {title} {\bibinfo {title} {{Renormalization-group
  approach to the Anderson model of dilute magnetic alloys. II. Static
  properties for the asymmetric case}},\ }\href
  {https://doi.org/10.1103/PhysRevB.21.1044} {\bibfield  {journal} {\bibinfo
  {journal} {Phys. Rev. B}\ }\textbf {\bibinfo {volume} {21}},\ \bibinfo
  {pages} {1044} (\bibinfo {year} {1980})}\BibitemShut {NoStop}%
\bibitem [{\citenamefont {Cao}\ \emph {et~al.}(2018{\natexlab{a}})\citenamefont
  {Cao}, \citenamefont {Fatemi}, \citenamefont {Demir}, \citenamefont {Fang},
  \citenamefont {Tomarken}, \citenamefont {Luo}, \citenamefont
  {Sanchez-Yamagishi}, \citenamefont {Watanabe}, \citenamefont {Taniguchi},
  \citenamefont {Kaxiras}, \citenamefont {Ashoori},\ and\ \citenamefont
  {Jarillo-Herrero}}]{TBG1}%
  \BibitemOpen
  \bibfield  {author} {\bibinfo {author} {\bibfnamefont {Y.}~\bibnamefont
  {Cao}}, \bibinfo {author} {\bibfnamefont {V.}~\bibnamefont {Fatemi}},
  \bibinfo {author} {\bibfnamefont {A.}~\bibnamefont {Demir}}, \bibinfo
  {author} {\bibfnamefont {S.}~\bibnamefont {Fang}}, \bibinfo {author}
  {\bibfnamefont {S.~L.}\ \bibnamefont {Tomarken}}, \bibinfo {author}
  {\bibfnamefont {J.~Y.}\ \bibnamefont {Luo}}, \bibinfo {author} {\bibfnamefont
  {J.~D.}\ \bibnamefont {Sanchez-Yamagishi}}, \bibinfo {author} {\bibfnamefont
  {K.}~\bibnamefont {Watanabe}}, \bibinfo {author} {\bibfnamefont
  {T.}~\bibnamefont {Taniguchi}}, \bibinfo {author} {\bibfnamefont
  {E.}~\bibnamefont {Kaxiras}}, \bibinfo {author} {\bibfnamefont {R.~C.}\
  \bibnamefont {Ashoori}},\ and\ \bibinfo {author} {\bibfnamefont
  {P.}~\bibnamefont {Jarillo-Herrero}},\ }\bibfield  {title} {\bibinfo {title}
  {{Correlated insulator behaviour at half-filling in magic-angle graphene
  superlattices}},\ }\href {https://doi.org/10.1038/nature26154} {\bibfield
  {journal} {\bibinfo  {journal} {Nature}\ }\textbf {\bibinfo {volume} {556}},\
  \bibinfo {pages} {80} (\bibinfo {year} {2018}{\natexlab{a}})}\BibitemShut
  {NoStop}%
\bibitem [{\citenamefont {Cao}\ \emph {et~al.}(2018{\natexlab{b}})\citenamefont
  {Cao}, \citenamefont {Fatemi}, \citenamefont {Fang}, \citenamefont
  {Watanabe}, \citenamefont {Taniguchi}, \citenamefont {Kaxiras},\ and\
  \citenamefont {Jarillo-Herrero}}]{TBG2}%
  \BibitemOpen
  \bibfield  {author} {\bibinfo {author} {\bibfnamefont {Y.}~\bibnamefont
  {Cao}}, \bibinfo {author} {\bibfnamefont {V.}~\bibnamefont {Fatemi}},
  \bibinfo {author} {\bibfnamefont {S.}~\bibnamefont {Fang}}, \bibinfo {author}
  {\bibfnamefont {K.}~\bibnamefont {Watanabe}}, \bibinfo {author}
  {\bibfnamefont {T.}~\bibnamefont {Taniguchi}}, \bibinfo {author}
  {\bibfnamefont {E.}~\bibnamefont {Kaxiras}},\ and\ \bibinfo {author}
  {\bibfnamefont {P.}~\bibnamefont {Jarillo-Herrero}},\ }\bibfield  {title}
  {\bibinfo {title} {{Unconventional superconductivity in magic-angle graphene
  superlattices}},\ }\href {https://doi.org/10.1038/nature26160} {\bibfield
  {journal} {\bibinfo  {journal} {Nature}\ }\textbf {\bibinfo {volume} {556}},\
  \bibinfo {pages} {43} (\bibinfo {year} {2018}{\natexlab{b}})}\BibitemShut
  {NoStop}%
\end{thebibliography}
\end{document}